\documentclass{elsarticle}
\usepackage{amsmath,bm,epsfig,subfigure}
\usepackage{epstopdf}
\usepackage[margin=0.6in]{geometry}
\usepackage{indentfirst}
\usepackage{amssymb}
\usepackage{graphicx}
\usepackage{float}
\usepackage{color}
\usepackage[normalem]{ulem}
\usepackage{lineno,hyperref}
\modulolinenumbers[5]
\usepackage[toc,page]{appendix}
\usepackage{natbib}
\usepackage{subfigure}

\journal{Philosophical Transactions of the Royal Society A,}
\usepackage{etoolbox}
\makeatletter
\patchcmd{\ps@pprintTitle}
  {Preprint submitted to}
  {(Published in}
  {}{}
\makeatother

\begin{document}

\title{Transient dynamics in strongly nonlinear systems: optimization of initial conditions on the resonant manifold}

\author{Nathan Perchikov}

\author[]{O.V. Gendelman\corref{mycorrespondingauthor}}
\cortext[mycorrespondingauthor]{Corresponding author}
\ead{ovgend@tx.technion.ac.il}

\address{Faculty of Mechanical Engineering, Technion, Haifa 32000, Israel}

\begin{abstract}
We  consider a system of two linear and linearly coupled oscillators with ideal impact constraints. Primary resonant energy exchange is investigated by analysis of the slow-flow using the action-angle (AA) formalism. Exact inversion of the action-energy dependence for the linear oscillator with impact constraints is not possible. This difficulty, typical for many models of nonlinear oscillators, is circumvented by matching the asymptotic expansions for the linear and impact limits. The obtained energy-action relation enables the complete analysis of the slow-flow and the accurate description of the critical delocalization transition. The transition from the localization regime to the energy-exchange regime is captured by prediction of the critical coupling value. Accurate prediction of the delocalization transition requires detailed account of the coupling energy with appropriate re-definition and optimization of the limiting phase trajectory on the resonant manifold.
\end{abstract}


\maketitle

\section{Introduction}

In the majority of typical designs in mechanical engineering and associated disciplines, dynamic elements work in the linear or the weakly nonlinear (quasilinear) regime. Such regimes are well-understood, predictable and assessable by well-developed methods of analysis \cite{NayfehMook,NayfehBala}. Those methods often rely on the ideas of averaging, multiple-scale expansions or other asymptotic techniques \cite{Arnold1989,Sanders2007,Awrejcewicz2012}. However, essential nonlinearities, both intentional and unintentional, do occur in mechanical systems, due to various reasons, related to clearances, impacts, friction, material nonlinearities and plasticity \cite{Fidlin2006,Babitsky1978,Pilipchuk2010}, to name but a few.

In many applications, such nonlinearities are crucial for the desired performance \cite{Fidlin2006,Babitsky1978}. In others cases, they are completely unwanted, but are nonetheless manifested in the dynamics, leading to profound implications. Cracks in continuous structures are one example of the phenomenon \cite{Babitsky2014,Hiwarkar2012,Andreaus2007}.

The analysis of the dynamics of essentially nonlinear systems is a major challenge. The full picture can only be obtained for the extremely rare completely integrable systems \cite{Arnold1989}. For other systems, it is sometimes possible to derive exact periodic solutions -- examples include nonlinear normal modes \cite{Rosenberg1962,Rand1974,Vakakis1996} and discrete breathers in selected models \cite{Ablowitz1976,Ovchinnikov1999,Gendelman2013}. Partial description of periodic solutions in a broad variety of systems can be obtained by approximate and numeric methods \cite{Peeters2009,Mikhlin1995,Sire2005}. However, in essentially nonlinear systems, for one, the superposition principle does not apply, and it is usually hardly possible to use it even approximately. Despite an abundance of valuable insights obtainable from periodic solutions, they are insufficient for the understanding of the transient dynamics and energy transport in essentially nonlinear systems. Yet the nonstationary processes are usually the most interesting and important ones for applications. A major progress in the theoretical study of energy transport in essentially nonlinear systems has been achieved since it was realized that the most efficient transport usually occurs under conditions of resonance. This observation permits one to treat the system in the vicinity of the resonance manifold (RM), and to restrict the consideration by addressing the averaged equations of motion (sometimes referred to as slow-flow equations). These crucial simplifications often result in the emergence of conservation laws absent in the full system outside the RM. In the particular case of a conservative system with two dynamic degrees of freedom, the existence of an additional integral of motion in the approximation with the isolated resonance leads to complete integrability. This classical result has been first formulated in the quasilinear setting and dates back to Birkhoff’s theory of normal forms \cite{Birkhoff1927,Moser1973,Verhulst1979,Augusteijn1980}. In \cite{Breitenberger1981}, the aforementioned approach was used for the exploration of beating in a spring pendulum under conditions of 1:2 resonance. A recent application \cite{Ianets2017} addresses the propagation of asymmetric Gaussian beams in nonlinear waveguides.

As mentioned above, the method of normal forms in its traditional setting requires the system to be quasilinear. Nevertheless, it is common to use formally similar methods for exploration of nonlinear systems far beyond the quasilinear regime. Harmonic balance with slowly varying amplitudes \cite{Hayashi2014} is an important example of this approach: it lacks rigorous mathematical justification, but often provides reliable results and is widely used in engineering. Expedient approximations, mathematically equivalent to the method of normal forms in the quasilinear case, can be obtained by invoking complex variables. Early examples of this approach are models that include self-trapping \cite{Eilbeck1985} and the rotating-wave approximation \cite{Flach2008}, common in studies of lattice dynamics. Recently, similar ideas were reformulated in the complexification-averaging (CxA) approach \cite{Manevitch2007,Manevitch2014,Manevitch2011,Kovaleva2016,ManevitchRomeo}. At the same time, it is evident that the applicability of harmonic-balance-based methods is limited. For starters, their validity can be rigorously justified only for quasilinear systems or for the case of nonlinear potentials with power close to 2 \cite{James2011}. Harmonic-balance-related methods are nevertheless also used for quartic or even stronger nonlinearities \cite{Yuli2011}, and do in fact yield reasonably good results, albeit with limited and \emph{a priori} unknown accuracy. Partial remedy for the case of periodic solutions can be achieved by accounting for multiple harmonics in the expansion. Still, the attempts to use slowly varying amplitudes for multiple harmonics may lead to a system that mathematically is even more complex than the initial one. Besides, for very interesting and important systems that include clearances, impacts or rotators, these methods have additional limitations. It is possible, however, to devise efficient approaches for the exploration of transient responses and targeted energy transfer in some systems that include a single vibro-impact element or a rotator \cite{GendelmanJSV2012,GendelmanSigalov2012}. While handling these models, one also invokes the exploration of RMs, but beyond the harmonic balance or CxA. The treatment heavily relies on particular simplifications available for impacts or for rotators.

It is well-known that resonant dynamics can be efficiently explored beyond the quasilinear approximation by employing action-angle (AA) variables \cite{Arnold1989,LandauLifshitz,GoldsteinPoole,Percival1987}. The AA variables were instrumental in formulation of many prominent results and theories. Among others, one can mention the theory of adiabatic invariants \cite{LandauLifshitz}, the formulation and proof of the KAM theorem \cite{Arnold1989,GoldsteinPoole,Percival1987}, the development of the canonical perturbation theory \cite{Arnold2006,Itin2007}, explorations on Hamiltonian chaos \cite{Zaslavsky2007,Chirikov1979}, autoresonant phenomena \cite{Fajans2001,Fajans1999}, targeted energy transfer \cite{Vakakis2001}, etc. The AA formalism has an important theoretical advantage: all the RMs can be described at the same level of complexity \cite{Sapsis2017}. Exploration of the dynamics on the RMs in terms of the AA variables yields all the regular benefits mentioned above (for example, the appearance of additional conservation laws). The aforementioned completely integrable 2DOF conservative system with the isolated resonance was conveniently reformulated in terms of AA variables and widely used in early works on resonant dynamics of low-DOF systems \cite{WalkerFord,Ford1970}. It is easy to demonstrate that the methods based on primary harmonic balance (rotating-wave, CxA) are particular cases of the AA-based approach, equivalent to the latter when it applies the canonical AA transformation to a linear oscillator \cite{Sapsis2017}.

Recently, the AA approach was used for the exploration of the resonant dynamics in a system of two coupled vibro-impact oscillators \cite{Sapsis2017} and that of a forced particle escaping from a potential well \cite{GendelmanEscape2017}. These works demonstrated the noteworthy efficiency of the method, extending far beyond the bounds of applicability of the method of harmonic balance, but also revealed some limitations.

First, as well known \cite{Percival1987}, the AA transformation can be performed in a tractable form only for a handful of model potentials. Second, in some typical-enough models, one encounters features that do not allow unambiguous treatment of the averaged model in the vicinity of the RM. The present paper explores the ideas that can help to overcome or circumvent these difficulties.

The AA formalism expresses the phase space coordinates in terms of angles, the time-derivatives of which are frequencies. For a system of two oscillators, the interesting non-trivial dynamics is related to energy exchange between the oscillators, which occurs through resonance corresponding to proximity of the frequencies, or the angles. Therefore, the AA formalism enables the direct addressing of the interesting dynamics of two-DOF systems. In the vicinity of 1:1 resonance conditions, the most efficient ones for energy transport, one can assume the proximity of the angles, which introduces a separation of time scales, related to oscillations represented by the small difference of the frequencies, and the finite sum of frequencies. The fast oscillations can then be averaged out to yield slow-flow equations. The averaging, along with conservation of energy renders the system essentially representable on a plane, where the global dynamics is regular (due to the appearance of the additional conservation law on the RM) and can be treated by traditional methods of qualitative analysis. The limit-phase trajectory (LPT), introduced by Manevitch \cite{Manevitch2014}, is a special constant-energy contour on the aforementioned plane, corresponding to initial conditions for which all the energy is placed in only one of the two oscillators. A critical situation corresponds to the LPT passing through a saddle of the averaged flow. It is shown in this work that in the case where the coupling energy depends on the actions of the oscillators, the LPT is defined ambigously. A generalization of the concept is suggested, which allows good prediction of critical transitions in systems suffering from the aforementioned ambiguity. Section \ref{Sec2} presents the model system using which the generalized concept is illustrated, in Sec. \ref{Sec3} the averaged system is derived, Sec. \ref{Sec4} gives asymptotic expansions for energy as a function of action for an anharmonic oscillator, necessary for the analysis, Sec. \ref{Sec5} exhibits the generalization of the concept of the LPT, in Sec. \ref{Sec6} the critical transition is obtained analytically for a limit case.  Section \ref{Sec7} presents numerical results for the general case and Sec. \ref{Sec8} contains the concluding remarks.

\section{Description of the model}
\label{Sec2}

\begin{figure}[!h]
\begin{center}
\includegraphics[scale =0.23,trim={0.05cm 0.7cm 0.5cm 0.25cm},clip]{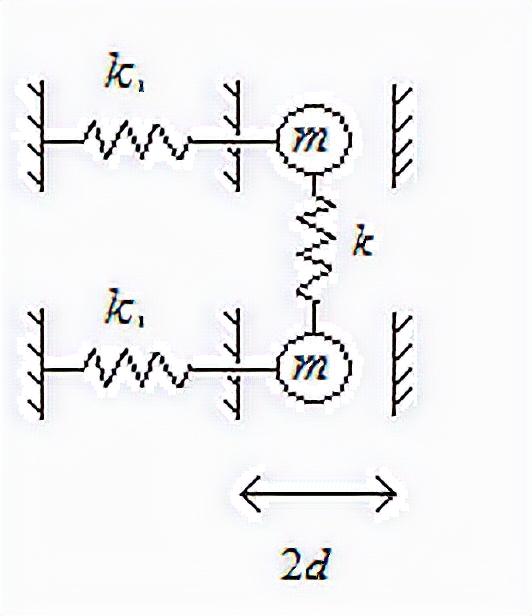}
\end{center}
\caption{\small A sketch of the analyzed system}
\label{Fig1}
\end{figure}

We consider two oscillators in on-site vibro-impact potential with harmonic smooth parts, connected by a linear spring, such that the motion is unidirectional. The on-site springs constant is denoted by $k_1$, the coupling (shear) spring constant is $k$, the identical masses are denoted by $m$, and the overall distance between impacts for each mass is $2d$. Ideal impacts are assumed. The system is depicted in Fig. \ref{Fig1}.

We follow the common approach, where for the sake of performing canonic AA transformations it is formally assumed that during a single fast-time oscillation cycle defining the action, there is no significant energy exchange between the oscillators. Following the understanding that efficient energy-exchange corresponds to resonance, the assumption is justified for the non-resonant regime. Then, the resonant manifold (RM) is derived for the main resonance. Assuming the transformation is justified even then, the canonic angles can then be related to actual frequencies, which are close at 1:1 resonance conditions, creating a slow energy-exchange time scale related to the frequency difference, for which the occurring energy exchange is efficient but slow and can be neglected for averaging over a fast oscillation period. This means that the transformation is self-consistent and qualitatively justified both on and off the RM. In line with this methodology, we will first treat a single oscillator, starting from Hamiltonian description, which reads:

\begin{equation}
\label{eq1}
H(p,q)=\frac{p^2}{2m}+\frac{1}{2}k_1 q^2=E
\end{equation}
where $E$ is the energy of a single oscillator, assumed to be conserved during a fast cycle (as aforementioned).

\section{Derivation of the slow-flow Hamiltonian}
\label{Sec3}

Next, the action and angle variables are introduced according to the definition
\begin{equation}
\label{eq2}
I(E)=\frac{1}{2\pi}\oint p(q,E) dq,\theta = \frac{\partial}{\partial{I}}\int_0^q p(\bar q,E)d\bar q
\end{equation}

Using dimensionless quantities, namely
\begin{equation}
\label{eq3}
\hat{E} \triangleq \frac{E}{\frac{1}{2}k_1d^2},\hat{I}\triangleq \frac{\omega_0 I}{\frac{1}{2}k_1d^2},\hat{q}\triangleq \frac{q}{d},\omega_0\triangleq\sqrt{\frac{k_1}{m}}
\end{equation}
one obtains from Eq. (\ref{eq1}) and the first of Eqs. (\ref{eq2}) the following relation:
\begin{equation}
\label{eq4}
\begin{split}
\hat{I}(\hat{E}) =\frac{2}{\pi}\left\lbrace\hat{E}\arcsin\left[\frac{1}{\sqrt{\max(1,\hat{E})}}\right]+\right.\left.\sqrt{\max(1,\hat{E})-1}\vphantom{\hat{E}\sin^{-1}\frac{1}{\sqrt{\max(1,\hat{E})}}}\right\rbrace
\end{split}
\end{equation}

Using the second of Eqs. (\ref{eq2}) and the derivative of Eq. (\ref{eq4}) and inverting to obtain a univalued function, one gets a formula for the coordinate:
\begin{equation}
\label{eq5}
\begin{split}
\hat{q}(\hat{E},\theta)   =  \sqrt{\hat{E}} \sin\left [\nu\arcsin(\sin\theta)\right]
\end{split}
\end{equation}
where
\begin{equation}
\label{eq8}
\begin{split}
\nu \triangleq \frac{2}{\pi}\arcsin\frac{1}{\sqrt{\max\left(1,\hat{E}\right)}} \le 1
\end{split}
\end{equation}

Next, following the methodology presented in \cite{Sapsis2017}, the expression given in Eq. (\ref{eq5}) is expanded into a $2\pi$-periodic Fourier sine series:
\begin{equation}
\label{eq6}
\begin{split}
\hat{q}(\hat{E},\theta) \sim \sum_{n=1}^{\infty}  A_n(\hat{E}) \sin(n\theta)
\end{split}
\end{equation}
where due to the equal frequencies of $\hat{q}$ and its expansion, one has
\begin{equation}
\label{eq7}
\begin{split}
 A_n(\hat{E}) =\frac{4}{\pi}\sqrt{\hat{E}} \int\limits_{0}^{\pi/2}  \sin\left (\nu\theta\right)\sin(n\theta)d\theta
=\frac{4}{\pi}\sqrt{\hat{E}}\frac{\nu\cos(\nu\pi/2)\sin(n\pi/2)}{n^2-\nu^2}
\end{split}
\end{equation}

At this point we pass to a system of \emph{two} oscillators of the type analyzed above. We assume that the energy exchange between the oscillators is either insignificant or significant but slow enough for the integral in Eq. (\ref{eq2}) to be performed under the assumption that the energy $E$ is constant during a fast oscillation cycle, yielding an adequate expression for the action, representative of the state of the oscillator, along with the angle of each oscillator. It will be shown later in this paper that for high enough coupling there can be fast significant energy exchange between the oscillators. In that case the transformation will not be formally justified. However, this does not pose a problem since the averaging procedure is only employed for the determination of the critical coupling and there is no intention to employ the transformation for finitely higher coupling values. 

Next, assuming linear coupling (with a linear or a shear spring represented by the coefficient $k$) between the oscillators, one can represent the total Hamiltonian of the system, using the two sets of action-angle variables, as follows:
\begin{equation}
\label{eq9}
\begin{split}
\hat H(\hat I_1,\hat I_2,\theta_1,\theta_2)=\hat E(\hat I_1)+\hat E(\hat I_2)
+\hat k\left[\hat{q}\left(\hat E(\hat I_1),\theta_1\right)-\hat{q}\left(\hat E(\hat I_2),\theta_2\right)\right]^2=\hat{E}
\end{split}
\end{equation}
where $\hat k \triangleq k/k_1,\hat H =H/(E/\hat E)$, and the function $\hat{E}(\hat{I})$ is the inverse function of the function $\hat{I}(\hat{E})$, given in Eq. (\ref{eq4}). We assume this inverse function exists and can be represented explicitly, at least approximately.

Substituting Eq. (\ref{eq6}) into Eq. (\ref{eq9}), one obtains the following form:
\begin{equation}
\label{eq10}
\begin{split}
\hat H(\hat I_1,\hat I_2,\theta_1,\theta_2)=\hat E(\hat I_1)+\hat E(\hat I_2)
+\hat k\left[\sum_{m,n=1}^{\infty}  A_m(\hat{E}(\hat{I}_1))A_n(\hat{E}(\hat{I}_1)) \psi_{mn}^{(1)}(\theta_1,\theta_2) \right. \\
+ \sum_{m,n=1}^{\infty}A_m(\hat{E}(\hat{I}_2))A_n(\hat{E}(\hat{I}_2))\psi_{mn}^{(2)}(\theta_1,\theta_2)
\left.-2 \sum_{m,n=1}^{\infty}A_m(\hat{E}(\hat{I}_1))A_n(\hat{E}(\hat{I}_2)) \psi_{mn}^{(3)}(\theta_1,\theta_2) \right]
\end{split}
\end{equation}
where
\begin{equation}
\label{eq11}
\begin{split}
\begin{gathered}[t]
\psi_{mn}^{(1)}(\theta_1,\theta_2) \triangleq  \sin(m\theta_1)\sin(n\theta_1) ,\
\psi_{mn}^{(2)}(\theta_1,\theta_2) \triangleq  \sin(m\theta_2)\sin(n\theta_2) ,\\
\psi_{mn}^{(3)}(\theta_1,\theta_2) \triangleq  \sin(m\theta_1)\sin(n\theta_2)
\end{gathered}
\end{split}
\end{equation}

Next we change the variables from independent angles to the sum and the difference of the angles,
\begin{equation}
\label{eq12}
\Theta \triangleq \theta_1+\theta_2 \ , \ \vartheta \triangleq \theta_1-\theta_2
\end{equation}

We seek to describe the nontrivial dynamics of the system, which, as aforementioned, occurs near the resonance, that is were the frequencies are close, or, in other words, were the difference of the frequencies is small relative to their sum (or any one angle, sense there is no other relevant magnitude reference). This yields the relation $\dot \vartheta \ll \dot \Theta$, which should, in principle, hold during most of the time taken by the dynamics we want to describe, and can thus be integrated to yield the relation $\vartheta \ll \Theta$. This means that there is a fast angle variable, $\Theta$, with respect to which the Hamiltonian can be averaged, to yield the slow flow. The dependence of the Hamiltonian on the \emph{angle} variables is given by the three functions in Eq. (\ref{eq11}, and hence, rewriting Eqs. (\ref{eq11}) in terms of the sum and the difference of the angles and averaging over the sum of the angles over a fast-oscillation period, the following averages are obtained:
\begin{equation}
\label{eq13}
\left\langle \psi_{mn}^{(1)}(\Theta,\vartheta)\right\rangle_{\Theta}=\left\langle \psi_{mn}^{(2)}(\Theta,\vartheta)\right\rangle_{\Theta}=\frac{1}{2}\delta_{mn} \ , \  \left\langle \psi_{mn}^{(3)}(\Theta,\vartheta)\right\rangle_{\Theta}=\frac{1}{2}\delta_{mn}\cos{(n\vartheta)}
\end{equation}
where $\delta_{mn}$ is Kronecker's delta.

By averaging Eq. (\ref{eq10}) over $\Theta$, employing Eqs. (\ref{eq12}) and (\ref{eq13}), introducing the slow-flow actions, defined as $\hat J_1\triangleq\langle\hat I_1\rangle_{\Theta},\hat J_2\triangleq\langle\hat I_2\rangle_{\Theta}$, and recalling the definition of the action as a quantity averaged over a fast oscillation cycle, which yields: $\langle\hat I_1\rangle_{\Theta}=\hat I_1,\langle\hat I_2\rangle_{\Theta}=\hat I_2$, we get the following averaged Hamiltonian:
\begin{equation}
\label{eq15}
\begin{split}
\left\langle\hat H\right\rangle_{\Theta}(\hat J_1,\hat J_2,\vartheta)=\hat E(\hat J_1)+\hat E(\hat J_2)+\\
+\frac{1}{2}\hat k\left[\sum_{n=1}^{\infty}  A_n^2(\hat{E}(\hat{J}_1))
+ \sum_{n=1}^{\infty}A_n^2(\hat{E}(\hat{J}_2))\right.
\left.- 2\sum_{n=1}^{\infty}A_n(\hat{E}(\hat{J}_1))A_n(\hat{E}(\hat{J}_2)) \cos{(n\vartheta)} \right]
\end{split}
\end{equation}

The dimensionality of the averaged system can be further reduced. Recalling that the AA variables are canonic, we can write
\begin{equation}
\label{eq16}
\begin{split}
\dot{\hat I}_1=-\frac{\partial\hat H}{\partial\theta_1},\dot{\hat I}_2=-\frac{\partial\hat H}{\partial\theta_2}
\end{split}
\end{equation}

Expanding, differentiating and averaging the functions in Eq. (\ref{eq11}) and substituting the results into Eqs.(\ref{eq10}) and (\ref{eq16}), and using the definition of the averaged actions, we get the following averages of the rates in Eqs. (\ref{eq16}):
\begin{equation}
\label{eq17}
\begin{split}
\dot{\hat J}_2=-\dot{\hat J}_1=\hat k\sum_{n=1}^{\infty}  A_n(\hat{E}(\hat{J}_1))A_n(\hat{E}(\hat{J}_2)) n\sin(n\vartheta)
\end{split}
\end{equation}

This yields an additional conservation law for the averaged flow on a RM, namely:
\begin{equation}
\label{eq18}
\begin{split}
\dot{\hat J}_1+\dot{\hat J}_2=0 \Rightarrow \hat{J}_1+\hat{J}_2=N^2=\text{const.}
\end{split}
\end{equation}
where $N$ can be termed the participation number (this result holds for 1:1 resonance, for higher-order resonances, there would be integer non-unity coefficients in the equation \cite{Sapsis2017}). This allows using the following trigonometric parametrization:
\begin{equation}
\label{eq19}
\begin{split}
\hat{J}_1=N^2\sin^2{\frac{\gamma}{2}}, \ \hat{J}_2=N^2\cos^2{\frac{\gamma}{2}},
\end{split}
\end{equation}
which reduces the averaged flow to the surface of a sphere, with $\vartheta \in [0,2\pi)$ (as the difference of two close angles should not be more than a cycle, for proper scale separation), and $\gamma \in [0,\pi]$ (since $\gamma/2$ in the first quarter-plane is sufficient for the conservation law to hold).

We thus obtain the following (conserved) reduced Hamiltonian for the system:
\begin{equation}
\label{eq20}
\begin{split}
h(\gamma,\vartheta)=\hat E\left(N^2\sin^2{\frac{\gamma}{2}}\right)+\hat E\left(N^2\cos^2{\frac{\gamma}{2}}\right)
+\frac{1}{2}\hat k\sum_{n=1}^{\infty}  A_n^2\left[\hat{E}\left(N^2\sin^2{\frac{\gamma}{2}}\right)\right] +\\
\frac{1}{2}\hat k\sum_{n=1}^{\infty}A_n^2\left[\hat{E}\left(N^2\cos^2{\frac{\gamma}{2}}\right)\right]
- \hat k\sum_{n=1}^{\infty} A_n\left[\hat{E}\left(N^2\sin^2{\frac{\gamma}{2}}\right)\right]  A_n\left[\hat{E}\left(N^2\cos^2{\frac{\gamma}{2}}\right)\right] \cos{(n\vartheta)} \\=\text{const.}
\end{split}
\end{equation}

A more explicit form can be obtained if one defines:
\begin{equation}
\label{eq21}
\begin{split}
\hat E_1(\gamma)\triangleq\hat E\left(N^2\sin^2{\frac{\gamma}{2}}\right), \
\nu_1(\gamma) \triangleq \frac{2}{\pi}\sin^{-1}\frac{1}{\sqrt{\max\left[1,\hat{E}\left(N^2\sin^2{\frac{\gamma}{2}}\right)\right]}}
\end{split}
\end{equation}
\begin{equation}
\label{eq22}
\begin{split}
\hat E_2(\gamma)\triangleq\hat E\left(N^2\cos^2{\frac{\gamma}{2}}\right), \
\nu_2(\gamma) \triangleq \frac{2}{\pi}\sin^{-1}\frac{1}{\sqrt{\max\left[1,\hat{E}\left(N^2\cos^2{\frac{\gamma}{2}}\right)\right]}}
\end{split}
\end{equation}

Consequently, the slow-flow Hamiltonian reads
\begin{equation}
\label{eq23}
\begin{split}
\begin{aligned}[t]
&h(\gamma,\vartheta)=\\&\hat E_1(\gamma)+\hat E_2(\gamma)
+\frac{8}{\pi^2}\hat k \sum_{n=1,3,5}^{\infty} \left\lbrace \hat E_1(\gamma)\frac{\nu_1^2(\gamma)\cos^2{\frac{\pi\nu_1(\gamma)}{2}}}{[n^2-\nu_1^2(\gamma)]^2} \right.  +\hat E_2(\gamma)\frac{\nu_2^2(\gamma)\cos^2{\frac{\pi\nu_2(\gamma)}{2}}}{[n^2-\nu_2(\gamma)^2]^2} \\
&
-2 \sqrt{\hat E_1(\gamma)\hat E_2(\gamma)}\frac{\nu_1(\gamma)\cos{\frac{\pi\nu_1(\gamma)}{2}}}{n^2-\nu_1(\gamma)^2}\left. \frac{\nu_2(\gamma)\cos{\frac{\pi\nu_2(\gamma)}{2}}}{n^2-\nu_2^2(\gamma)} \cos{(n\vartheta)} \right\rbrace =\text{const.}
\end{aligned}
\end{split}
\end{equation}

Aside from the issue of initial conditions (and their relation to the parameter $N$), which would be addressed at a later stage, what is left to do in order to be able to describe the averaged flow completely, is to obtain the inversion of the function given in Eq. (\ref{eq4}) -- that is to obtain $\hat{E}(\hat J)$.

For the case where the energy of the oscillator is relatively small, the solution for a simple harmonic oscillator is reproduced, which in the normalized quantities gives the following (exact) relation:
\begin{equation}
\label{eq24}
\hat{E}(\hat{J})= \hat{J} \ , \ \forall \ \hat J \in [0,1]
\end{equation}

\section{Asymptotic expansion of the energy-action mapping for the nonlinear regime}
\label{Sec4}

\subsection{The small impact-velocity limit}
\begin{figure}[!h]
\begin{center}
{\includegraphics[scale = 0.45]{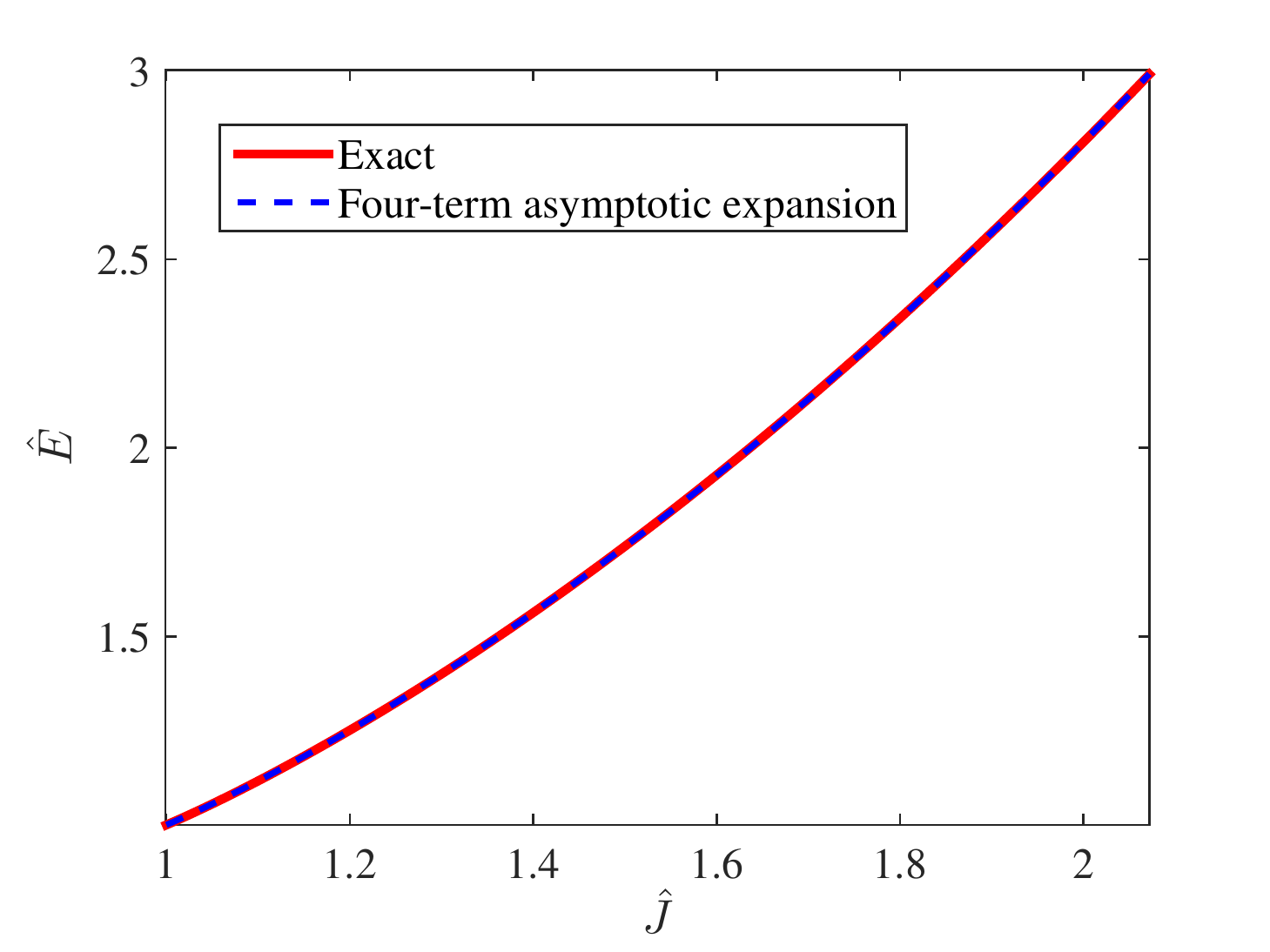}
\includegraphics[scale = 0.45]{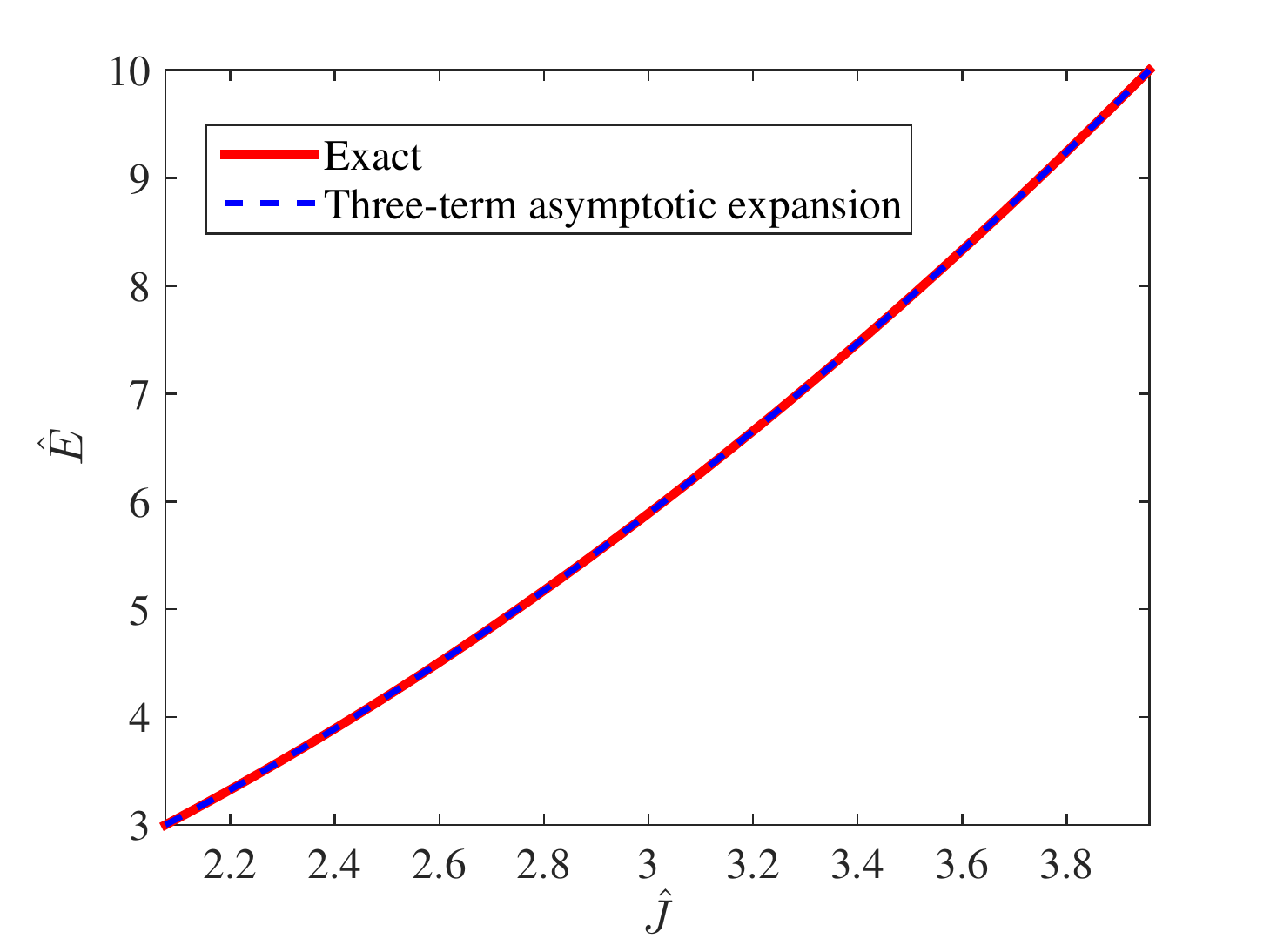}}
\end{center}
\caption{\small Comparison of the exact energy-action mapping (Eq. (\ref{eq4})) and a four-term asymptotic expansion (Eq. (\ref{eq27})), for $\hat{E}\in[1,3]$ (left) and the exact energy-action mapping (Eq. (\ref{eq4})) and a three-term asymptotic expansion (Eq. (\ref{eq30})), for $\hat{E}\in (3,10]$ (right)}
\label{Fig2}
\end{figure}

For the small impact-velocity limit, a sufficient approximation can be obtained by  using the identity: $\sin^{-1}\hat{E}^{-1/2}=\pi/2-\sin^{-1}\sqrt{1-\hat{E}^{-1}}$, expanding the small-argument arcsine to get a four-term expansion, namely,
\begin{equation}
\label{eq25}
\begin{split}
\sin^{-1}\frac{1}{\sqrt{\hat{E}}}\underset{\hat{E}\to 1^+}\to \frac{\pi}{2}-\sqrt{1-\frac{1}{\hat{E}}}-\frac{1}{6}\left(1-\frac{1}{\hat{E}}\right)^{3/2}-\frac{3}{40}\left(1-\frac{1}{\hat{E}}\right)^{5/2}
\end{split}
\end{equation}
substituting in Eq. (\ref{eq4}), and again retaining four terms, which yields:
\begin{equation}
\label{eq26}
\hat{J}(\lambda)\underset{\lambda\to 0^+}\to 1+\lambda^2-\frac{4}{3\pi}\lambda^3+\frac{4}{15\pi}\lambda^5
\end{equation}
where $\lambda \triangleq \sqrt{\hat{E}-1} \ll 1$. Next, the expansion in Eq. (\ref{eq26}) is inverted asymptotically to produce the following consistent four-term expansion:
\begin{equation}
\begin{split}
\label{eq27}
\hat{E}(\hat{J})\underset{\hat{J}\to 1^+}\to \hat{E}^-(\hat{J})\triangleq\hat{J}+\frac{4}{3\pi}(\hat{J}-1)^{3/2}+\frac{8}{3\pi^2}(\hat{J}-1)^2 +\frac{840-36\pi^2}{135\pi^3}(\hat{J}-1)^{5/2}
\end{split}
\end{equation}
Eq. (\ref{eq27}) is in excellent agreement with Eq. (\ref{eq4}) for impact energies up to twice as large as the maximum potential energy, as shown in Fig. \ref{Fig2}.

Eqs. (\ref{eq24}) and (\ref{eq27}) can be used to analyze the dynamics of a system of two oscillators one of which has no impacts at all and the other has weak conservative impacts (the localization regime).

\subsection{The high impact-velocity limit}

Using the three-term asymptotic expansion for the arcsine, similar to what was used above, only for high energies, namely,
\begin{equation}
\label{eq28}
\begin{split}
\sin^{-1}\frac{1}{\sqrt{\hat{E}}}\underset{\hat{E}\gg 1}\to \hat{E}^{-\frac{1}{2}}+\frac{1}{6}\hat{E}^{-\frac{3}{2}}+\frac{3}{40}\hat{E}^{-\frac{5}{2}}
\end{split}
\end{equation}
and employing it for the asymptotic expansion of Eq. (\ref{eq4}), the following three-term expansion is obtained:
\begin{equation}
\label{eq29}
\begin{split}
\hat{J}(\hat{E})\underset{\hat{E}\gg 1}\to  \frac{4}{\pi}\hat{E}^{\frac{1}{2}}-\frac{2}{3\pi}\hat{E}^{-\frac{1}{2}}-\frac{1}{10\pi}\hat{E}^{-\frac{3}{2}}
\end{split}
\end{equation}

Eq. (\ref{eq29}) can then be asymptotically inverted to yield the following three-term expansion:
\begin{equation}
\begin{split}
\label{eq30}
\hat{E}(\hat{J})\underset{\hat{J} \gg 1}\to \hat{E}^+(\hat{J}) \triangleq \frac{\pi^2}{16}\hat{J}^2+\frac{1}{3}+\frac{16}{45\pi^2}\hat{J}^{-2}
\end{split}
\end{equation}

Eq. (\ref{eq30}) is in excellent agreement with Eq. (\ref{eq4}) for impact energies even as small as twice  the maximum potential energy, as shown in Fig. \ref{Fig2} (and obviously reproduces it in the high-energy limit).

Eqs. (\ref{eq24}) and (\ref{eq30}) can be used to analyze the dynamics of a system of two oscillators one of which has no impacts at all and the other has strong conservative impacts (the localization regime).

The three expressions given by Eqs. (\ref{eq24}), (\ref{eq27}) and (\ref{eq30}), can be used to construct a function defined by cases, which can then be used for numerical analysis of the averaged flow described by the Hamiltonian in Eq. (\ref{eq23}).

\subsection{Matched asymptotics}

In order to construct a function using the limiting cases for the overall description of the energy-action relation, the most natural approach is to choose a proper patching point. To this end it is beneficial to define the mutual error function as the square root of the sum of the squares of the errors of the two functions, and to find its minimum, which would correspond to the point where both asymptotic expansion work reasonably well. This point can be chosen as the patching point. We define
\begin{equation}
\begin{split}
\label{eq31}
\mathcal{E} (\hat{E})\triangleq \left ( \left\lbrace\hat{E}^+[\hat{J}(\hat{E})]-\hat{E}  \right \rbrace^2+\right. \left.\left\lbrace \hat{E}^-[\hat{J}(\hat{E})]-\hat{E}\right\rbrace^2\right ) ^{1/2}
\end{split}
\end{equation}
where $\hat{J}(\hat{E})$ refers to the exact function given in Eq. (\ref{eq4}). The mutual error function can be obtained by choosing a proper range for $\hat{E}$, calculating $\hat{J}(\hat{E})$ from Eq. (\ref{eq4}), and then using the approximations in Eqs.  (\ref{eq27}) and (\ref{eq30}). Clearly, the least-mutual-error point should neither be too close to nor too far from unity. As suggested by Fig. \ref{Fig2}, a good estimate would be $\hat{E}=3$.

\begin{figure}[!h]
\begin{center}
\includegraphics[scale = 0.45]{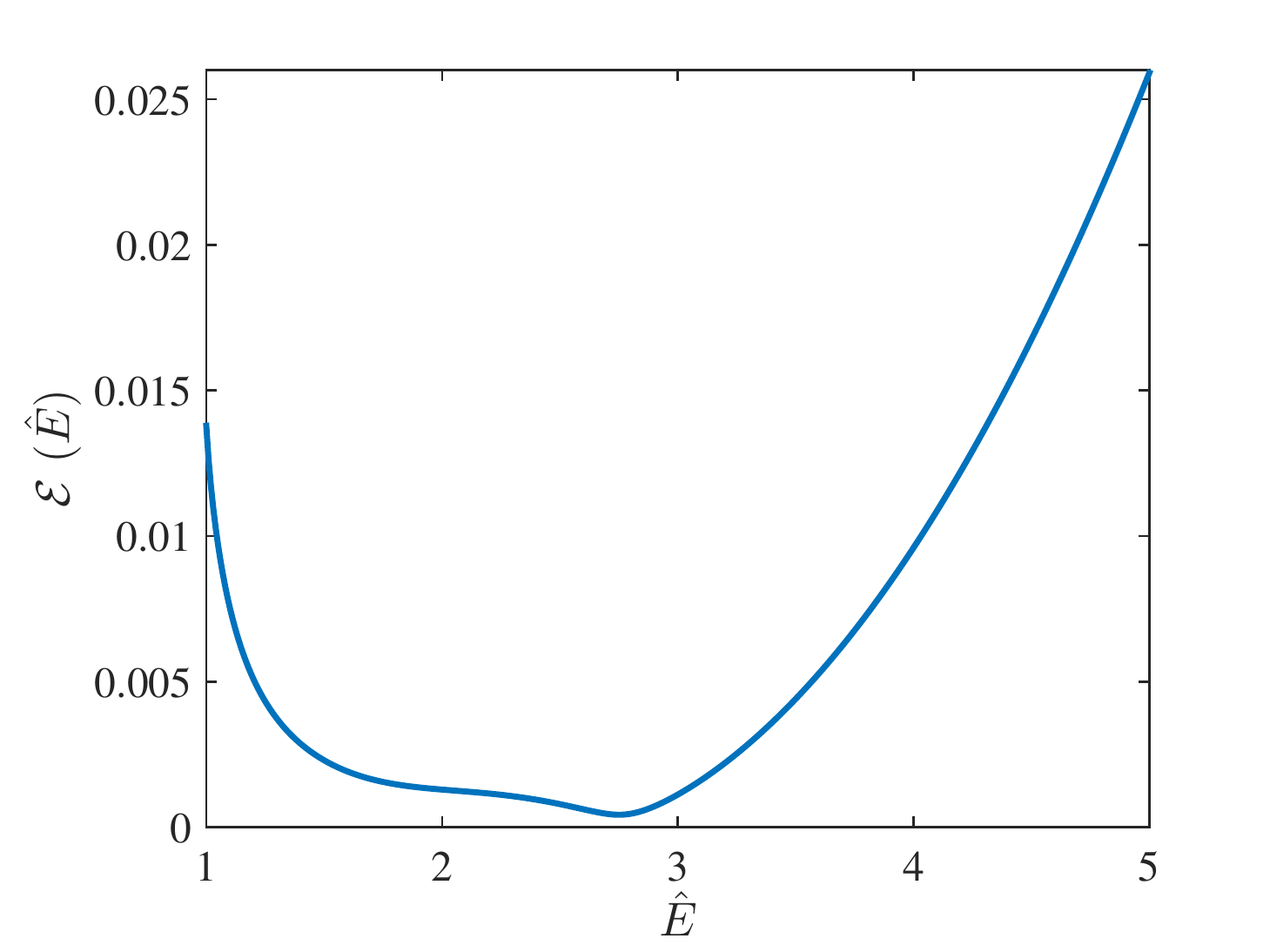}
\end{center}
\caption{\small Mutual error of the asymptotic expansions}
\label{Fig4}
\end{figure}

The curve in Fig. \ref{Fig4} has a minimum at $\hat{E}^* \approx 2.75$. The corresponding value of the action is $\hat{J}^* \approx 1.975$. The value of the relative minimum mutual error is $\mathcal{E}(\hat{E}^*)/\hat{E}^* \approx 1.55 \times 10^{-4}$.

If this point is used as the patching point, or the switch from $\hat{E}^-(\hat{J})$ to $\hat{E}^+(\hat{J})$, then one has $|\hat{E}^-(\hat{J})-\hat{E}|<10^{-3}$ and $|\hat{E}^+(\hat{J})-\hat{E}|<5\times 10^{-4}$, which means that if we define:
\begin{equation}
\begin{split}
\label{eq32}
\hat{E}^{(-/+)}(\hat{J})\triangleq \begin{cases} \hat{J} \ , \ \forall \ \hat{J} \in [0,1] \\ \hat{E}^-(\hat{J}) \ , \ \forall \ \hat{J}\in [1,\hat{J}^*]  \\  \hat{E}^+(\hat{J}) \ , \ \forall \ \hat{J}>\hat{J}^* \end{cases}
\end{split}
\end{equation}
then one has $|\hat{E}^{(-/+)}[\hat{J}(\hat{E})]-\hat{E}|/\hat{E}<10^{-3} \ \forall \ \hat{E}$.

Therefore it can be concluded that $\hat{E}^{(-/+)}$ should be sufficient for the approximate analysis.

It can happen that two asymptotic expansions corresponding to opposite limits never intersect. In such cases, using the least mutual error for the patching point is a possible alternative. Still, having a continuous function adequate in the full range is advantageous. In the considered case, there are two points where the functions $\hat{E}^{(-)}$ and  $\hat{E}^{(+)}$ intersect. One point is around $\hat{J}=1.67$. The second intersection point is closer to the point of minimal mutual error and thus it is globally a better choice. The point is $\hat{J}_p = 1.8829579828475$, and if used instead of $\hat{J}^*$ in Eq. (\ref{eq32}), it produces a continuous approximation of $\hat{E}(\hat{J})$, with a maximum relative error of $|\hat{J}_p-\hat{I}(\hat{E}^{(-/+)}(\hat{J}_p))|/\hat{I}(\hat{E}^{(-/+)}(\hat{J}_p)))<4.3\times10^{-4}$, where $\hat{I}(\hat{E}^{(-/+)})$ is taken from Eq.(\ref{eq4}). Both Eq. (\ref{eq32}) and the continuous version with $\hat{J}_p$ were used in this work, producing indistinguishable results, with the continuous version used for the results presented in the numerical results section. For calculations requiring the derivative of $\hat{E}^{(-/+)}(\hat{J})$, one can obtain a continuous derivative using the quantity $1/\hat{I}'(\hat{E}^{(-/+)}(\hat{J}))$. The latter was done in the present work for estimation of the frequencies, which in the canonic formalism are given by $\omega_{1,2}=\partial{h}/\partial{\hat{J}}_{1,2}$ (not strictly required for the critical transition analysis and used only for consistency validation).

\section{Generalization of the concept of the Limiting Phase Trajectories (LPT)}
\label{Sec5}

\subsection{Initial conditions for the LPT -- the general idea}

The limiting phase trajectory (LPT) is a concept formally introduced by Manevitch \cite{Manevitch2014}. It is usually defined for the slow-flow of a specific dynamical system and refers to a specific trajectory that passes through the point corresponding to localized initial conditions. For a system of two dynamic degrees of freedom, the LPT (ideally) passes through a state where all the energy is contained in the initial momentum of one of the particles. The benefit of using this trajectory is that it is helpful in determining the transition between a regime of dynamic localization to the regime of energy exchange between the physical oscillators comprising the system. For parameter values for which the LPT passes through a saddle point in RM of the slow-flow, one should expect the original system to show a transition from the localization regime to the regime of intensive energy exchange.

One problem related to the approach of the LPT-saddle point coincidence is that in many scenarios the initial condition of complete localization of the full system is not represented in the averaged system. This can be illustrated on the example of the system of two oscillators in two-sided on-site vibro-impact potential with linear coupling. For the case of no foundation (that is only impact constraints, with no linear on-site springs), as shown in \cite{Sapsis2017}, the averaged Hamiltonian only depends on $\gamma$ through the expressions for the independent energies of the oscillators, $\hat{E}_{1,2}$. Therefore, exact (initial) localization, say, on the second particle, is represented by the condition $\gamma=0$. The energy contained in the coupling potential is nonzero on average for the initial value of the slow phase $\vartheta$ (which can be set to zero, for the estimation of the initial condition, with no loss of generality), however it is independent of $\gamma$, and therefore raises no question regarding the correct definition of the initial value of $\gamma$ that corresponds to the full localization.

For the case of nonzero linear on-site springs, however, the coupling-energy part of the averaged Hamiltonian does depend on $\gamma$, and it becomes unclear that indeed the choice $\gamma=0$ represents utmost initial localization (corresponding to the energy being localized in the momentum of one particle).

This observation is exhibited for the considered model by direct calculation. It is shown that taking the $\gamma=0$ as the initial condition for the LPT in the limit of infinitely weak on-site linear springs, leads to a finite non-negligible discrepancy with the case where there are no linear on-site springs per se, a discrepancy evident for example in the important question of the critical coupling for slow-flow delocalization.

The natural remedy for the aforementioned problem is to define the initial condition for the action-parameterizing angle $\gamma$ by the condition that it should maximize the level of localization of the energy in the momentum of one (for instance, the second) particle, or (maximize) the kinetic energy of the second oscillator relative to the total energy of the system, averaged over the fast angle. Thus the problem is formulated as a maximization problem with respect to the argument $\gamma$.

Clearly, from a qualitative perspective, the notion of localization generally requires that one (say the first) oscillator initially does not hit the walls, experiencing only linear \emph{individual} dynamics.

\subsection{Initial conditions for the LPT -- specific considerations}

The general idea of the LPT is that it should describe, on average, the transition of the full system from a state where all the energy is contained in the momentum of one oscillator (only at the initial instance of a fast oscillation cycle) to a state of primary resonance with equal distribution of energy between the oscillators. As aforementioned, after averaging over the fast oscillations, the minimum requirement for correct representation of the full system is that one oscillator is not impacting during several initial fast cycles. The second oscillator does impact and the time between impacts is related to the fast oscillation frequency. One therefore seeks an expression for the Hamiltonian averaged over that fast angle, say, $\theta_2$, which could be used for the determination of the initial condition for the LPT.
An important problem in the LPT-based approach, complementing the problem noted in the previous subsection (related to the proper value of $\gamma_0$), is that resonant conditions in the full system generally do not start immediately. First, there is a period of time where the phases of the oscillators are coordinated in a certain manner and only then does initial small energy exchange begin eventually emerging as resonance. Therefore, on cannot in principle assume that in the state of utmost localization the system, averaged over the fast angle, already lies on the RM. Consequently, one cannot simply assumed that Eq.(\ref{eq23}) is valid at the initial condition for LPT and search for the right $\gamma_0$ value. Therefore, in order to determined the initial conditions for the LPT, one must start with  the general expression for the Hamiltonian given in Eq. (\ref{eq9}), averaged over the fast angle, $\theta_2$:
\begin{equation}
\label{eqIC1}
\begin{split}
\langle\hat H\rangle_{\theta_2}(\hat J_1,\hat J_2,\theta_1(t))=\hat E(\hat J_1)+\hat E(\hat J_2)
+\hat k\left[\langle\hat{q}_1^2\rangle_{\theta_2}+\langle\hat{q}_2^2\rangle_{\theta_2}-2\langle\hat{q}_1\hat{q}_2\rangle_{\theta_2}\right]
\end{split}
\end{equation}
(where the actions and energies are angle-independent in the canonic AA formalism, and thus $\hat{I}$ was replaced by $\hat{J}$ everywhere, and it does not matter whether the averaging is with $\theta_2$ or with $\Theta$). The expression in Eq. (\ref{eqIC1} is correct even outside the RM. The next step is to use the aforementioned assumption that the first oscillator does not impact during the initial fast cycles. Without impacts, and considering that energy exchange has not yet began at the sought point, one therefore assumes linear dynamics for the first oscillator, for which one has: $\hat{E}(\hat{J}_1)=\hat{J}_1,\hat{q}_1=\sqrt{\hat{J}_1}\sin{\theta_1(t)}$, and hence
\begin{equation}
\label{eqIC2}
\begin{split}
\langle\hat H\rangle_{\theta_2}(\hat J_1,\hat J_2,\theta_1(t))_{\text{IC}}=\hat J_1+\hat E(\hat J_2)
+\hat k\left[\hat{J}_1\langle\sin^2{\theta_1(t)}\rangle_{\theta_2}+\langle\hat{q}_2^2\rangle_{\theta_2}-2\sqrt{\hat{J}_1}\langle\sin{\theta_1(t)}\hat{q}_2\rangle_{\theta_2}\right]
\end{split}
\end{equation}

What we need for an initial condition (IC) for an LPT is the best representation of the full system, in which the energy not contained in the second oscillator's momentum is minimal initially. Therefore for utmost representation of the full system in the IC for the LPT, we need to minimize the first and third terms in Eq. (\ref{eqIC2}) with respect to the variables that remain free. We have averaged with respect to $\theta_2$, and $\hat{J}_2$ is related to the conserved overall energy, and thus we need to minimize with respect to $\hat{J}_1$ and $\theta_1(t)$. The first step is to minimize with respect to $\theta_1$.

Now, the only term that depends on $\theta_1$ is the term in the brackets, that multiplies the coupling stiffness. Therefore, we need to find its minimum. Moreover, since the term in the brackets is averaged in effect over time, it is not a function of some initial value of $\theta_1$, but rather a functional depending on the function $\theta_1(t)$ in a certain range, representing several initial fast oscillations. Therefore, we seek a variational minimum of the term in the brackets in Eq. (\ref{eqIC2}).

To find the aforementioned variational minimum may be a complicated task. Although the unconstrained minimum is $\hat{q}_1(t)=\hat{q}_2(t)$, it is infeasible, since it violates the assumption $\hat{J}_1<1$ for $\hat{J}_2>1$. So instead, we will find the variational \emph{infimum} of the expression. Following the Cauchy-Schwartz inequality, one can write $\langle\hat{q}_1\hat{q}_2\rangle_{\theta_2}\le \sqrt{\langle\hat{q}_1^2\rangle_{\theta_2}\langle\hat{q}_2^2\rangle_{\theta_2}}$, which produces the infimum for Eq. (\ref{eqIC2}), as follows:
\begin{equation}
\label{eqIC3}
\begin{split}
\langle\hat H\rangle_{\theta_2}(\hat J_1,\hat J_2,\theta_1(t))^{\text{inf}}_{\text{IC}}=\hat J_1+\hat E(\hat J_2) + \\
+\hat k\left[\hat{J}_1\langle\sin^2{\theta_1(t)}\rangle_{\theta_2}+\langle\hat{q}_2^2\rangle_{\theta_2}-2\sqrt{\hat{J}_1}\sqrt{\langle\sin^2{\theta_1(t)}\rangle_{\theta_2}\langle\hat{q}_2^2\rangle_{\theta_2}}\right]
\end{split}
\end{equation}

The next step is the actual act of averaging with respect to $\theta_2$. The value of $\langle\sin^2{\theta_1(t)}\rangle_{\theta_2}$ depends on the order of $\dot\theta_1/\dot\theta_2$. If $|\dot\theta_1|\ll|\dot\theta_2|$, then the coupling energy, as can be learned already from Eq. (\ref{eqIC2}), is of the order of $\mathcal{O}(\hat{k})$ (as $\hat{J}_1<1$). If one considers the critical case of delocalization transition, then the coupling energy in the state assumed here to be that of localization becomes $\mathcal{O}(\hat{k}_{\text{cr}})$. As we will see in the following, and as one can guess, the coupling energy in the beating or energy exchange regime is of the order of the total energy. This means that if indeed $|\dot\theta_1|\ll|\dot\theta_2|$, then the coupling energy contains a significant portion of the total energy, which is not appropriate for the initial condition of localization (this argumentation also assumes the trivial fact that $\langle\hat{q}_2^2\rangle_{\theta_2}=\mathcal{O}(1)$, which is obvious for the normalized displacement square average of an impacting mass).

The conclusion from this lengthy argumentation is that the assumption $|\dot\theta_1|\ll|\dot\theta_2|$ is contradictory for (the perfectly viable) initial conditions of localization. Consequently, it is either that  $|\dot\theta_1|=\mathcal{O}(|\dot\theta_2|)$ or that  $|\dot\theta_1|\gg|\dot\theta_2|$ (in the initial localization regime). After that fact is established, one observes that for the case $|\dot\theta_1|\gg|\dot\theta_2|$, clearly one has $\langle\sin^2{\theta_1(t)}\rangle_{\theta_2}=1/2$. Furthermore, since the averaging with respect to $\theta_2$ can always be done not on one but on several of the first fast (second mass impact-) cycles, clearly for the case $|\dot\theta_1|=\mathcal{O}(|\dot\theta_2|)$, one would also have $\langle\sin^2{\theta_1(t)}\rangle_{\theta_2}=1/2$. A typical example is $|\dot\theta_1|=(|\dot\theta_2|/2)$, for which even for one cycle $\langle\sin^2{\theta_1(t)}\rangle_{\theta_2}=1/2$ (due to the square). If the ratio is slightly different, several cycles may be used for averaging. A problem would arise only for a small value of the aforementioned ratio, since then too many cycles may be needed, too much for example for the AA transformation to still be valid.

This way we established that for consistent  initial conditions of localization, where the only assumption is a \emph{qualitative one}, namely that there is localization and that there is consistency in the procedure, and using only orders-of-magnitude considerations, we derived a \emph{quantitative} result, as follows:
\begin{equation}
\label{eqIC4}
\begin{split}
\langle\hat H\rangle_{\theta_2}(\hat J_1,\hat J_2,\theta_1(t))^{\text{inf}}_{\text{IC}}=\hat J_1+\hat E(\hat J_2)
+\hat k\left[\langle\hat{q}_2^2\rangle_{\theta_2}+\frac{1}{2}\hat{J}_1-\sqrt{2\langle\hat{q}_2^2\rangle_{\theta_2}}\sqrt{\hat{J}_1}\right]
\end{split}
\end{equation}
(to be more precise, using \emph{weak} restrictions we obtained the above result, which is \emph{strong}, as will be shown next).

Next we present an expression for the Hamiltonian under the assumption that the first mass does not impact, and the frequencies are such that the dynamics of the slow-flow lies on the RM, taking (with no loss of generality) the initial angle difference of $\vartheta=0$:
\begin{equation}
\label{eq23w}
\begin{split}
h(\gamma,0)=\hat J_1(\gamma)+\hat E(\hat J_2(\gamma)) + \\
+\hat k  \left\lbrace \frac{8}{\pi^2}\sum_{n=1,3,5}^{\infty}\frac{\nu_2^2(\gamma)\cot^2{\frac{\pi\nu_2(\gamma)}{2}}}{[n^2-\nu_2(\gamma)^2]^2} +\frac{1}{2}\hat J_1(\gamma)
- \frac{4}{\pi}{{}}{}  \frac{\nu_2(\gamma)\cot{\frac{\pi\nu_2(\gamma)}{2}}}{1-\nu_2^2(\gamma)}\sqrt{\hat J_1(\gamma)}  \right\rbrace
\end{split}
\end{equation}
where one identifies the first term in the braces in Eq. (\ref{eq23w}) as $\langle\hat{q}_2^2\rangle_{\theta_2}$.

Now, the conceptual difference between the averaged Hamiltonians in Eqs. (\ref{eqIC4}) and (\ref{eq23w}) is that in terms of the function $\theta_1(t)$, the first corresponds to the limit of localization, and the second to the limit of delocalization. Thus the two constitute the two limit cases and given also the fact that the first is an infimum, one can argue that the true Hamiltonian for localization conditions should be bounded by the two expressions. In terms of form, the only difference between the two Hamiltonians is in the prefactor of $\sqrt{\hat{J}_1}$. We will denote this prefactor (without the minus) by $\xi$. One therefore has
\begin{equation}
\label{eqIC5}
\xi^{\text{inf}}_{\text{IC}}=\sqrt{2\langle\hat{q}_2^2\rangle_{\theta_2}} , \ \xi_{\text{RM}}= \frac{4}{\pi}{{}}{}  \frac{\nu_2(\gamma)\cot{\frac{\pi\nu_2(\gamma)}{2}}}{1-\nu_2^2(\gamma)}
\end{equation}

And the true Hamiltonian corresponding to the initial conditions of localization would be
\begin{equation}
\label{eqIC6}
\begin{split}
\langle\hat H\rangle_{\theta_2}(\hat J_1,\hat J_2,\theta_1(t))_{\text{IC}}=\hat J_1+\hat E(\hat J_2)
+\hat k\left[\langle\hat{q}_2^2\rangle_{\theta_2}+\frac{1}{2}\hat{J}_1-\xi\sqrt{\hat{J}_1}\right]
\end{split}
\end{equation}
with $\xi \in (\xi_{\text{RM}},\xi^{\text{inf}}_{\text{IC}}]$ (the order is determined by the fact that $-\xi^{\text{inf}}_{\text{IC}}$ produces the infimum so it is smaller and its negative is larger).

The main premise of this derivation is that if the values of $\xi_{\text{RM}}$ and $\xi^{\text{inf}}_{\text{IC}}$ are close enough, then the true Hamiltonian at localization initial conditions and an expression obtained under the assumptions of resonance, are close enough such that any of the expressions could be used for the construction of the LPT. If indeed this is the case, then considerations of convenience should determine the choice. Clearly then it would be advantageous to use the RM Hamiltonian, since the point of maximum energy exchange (the saddle point of the slow-flow, as will be shown next) lies on the RM and thus it would be possible to construct the entire trajectory on the RM. Even more importantly, since the quantity that remains conserved on the slow flow RM is $N$, and the quantity that one would want to control is the energy $E$, clearly using the RM Hamiltonian would be advantageous since then the initial action $\hat{J}_2$ could be expressed through $N$ and related to the energy. However, such advantageous a choice would only be justified if $\xi_{\text{RM}}$ and $\xi^{\text{inf}}_{\text{IC}}$ are close enough. It is shown in the following sections that this is indeed the case.

Now, once the issue of the choice of the correct value of $\xi$ has been cleared, there remains the issue of the choice of the action $\hat{J}_1$ corresponding to initial conditions of localization. Here the argumentation is more straightforward. If we use the Hamiltonian in Eq. (\ref{eqIC6}) and the issue of the choice of the value of $\xi$ is resolved (say, by the \emph{justified} choice $\xi=\xi_{\text{RM}}$), then simply in order to \emph{better represent} the initial conditions of the full system, one should minimize the part of the energy that is not contained in the average kinetic energy of the second oscillator, namely, by taking
\begin{equation}
\begin{split}
\label{eq33}
\begin{aligned}[t]
      &\gamma_0= \underset{\gamma}{\text{argmax}} \frac{\hat{E}_2(\gamma) - \langle\hat{q}_2^2\rangle_{_{\Theta}}(\gamma)}{\hat{E}_1(\gamma) +\hat{E}_2(\gamma)+\hat{k} \langle(\hat{q}_2-\hat{q}_1)^2\rangle_{_{\Theta}}(\gamma)}\\
      &s.t. \ \ \ \hat{E}_1(\gamma)<1,  N=\text{const.}
\end{aligned}
\end{split}
\end{equation}
where employing Eq. (\ref{eq23}), one obtains:

\begin{equation}
\begin{split}
\label{eq34}
\begin{aligned}[t]
	&\gamma_0= \underset{\gamma}{\text{argmax}} \left ( \left [ \vphantom{ \sum_{n=1,3,5}^{\infty}\frac{\nu_1^2(\gamma)\cot^2{\frac{\pi\nu_1(\gamma)}{2}}}{[n^2-\nu_1^2(\gamma)]^2}}\hat E_1(\gamma)+\hat E_2(\gamma) \right. \right.
+\frac{8}{\pi^2}\hat k \sum_{n=1,3,5}^{\infty} \left\lbrace\hat E_1(\gamma)\frac{\nu_1^2(\gamma)\cos^2{\frac{\pi\nu_1(\gamma)}{2}}}{[n^2-\nu_1^2(\gamma)]^2} +  \right. \\
&
+ \hat E_2(\gamma)\frac{\nu_2^2(\gamma)\cos^2{\frac{\pi\nu_2(\gamma)}{2}}}{[n^2-\nu_2(\gamma)^2]^2} -
2 \sqrt{\hat E_1(\gamma)\hat E_2(\gamma)}\frac{\nu_1(\gamma)\cos{\frac{\pi\nu_1(\gamma)}{2}}}{n^2-\nu_1(\gamma)^2} \left. \left. \frac{\nu_2(\gamma)\cos{\frac{\pi\nu_2(\gamma)}{2}}}{n^2-\nu_2^2(\gamma)}  \right\rbrace  \right ]^{-1} \\
 &
\left. \times E_2(\gamma) \left\lbrace 1 - \frac{8}{\pi^2} \sum_{n=1,3,5}^{\infty} \frac{\nu_2^2(\gamma)\cos^2{\frac{\pi\nu_2(\gamma)}{2}}}{\left[n^2-\nu_2^2(\gamma)\right]^2} \right\rbrace \right )\\
 &s.t. \ \ \ \hat{E}_1(\gamma)<1,  N=\text{const.}
\end{aligned}
\end{split}
\end{equation}

This is the maximization of the ratio of the kinetic energy of the second oscillator to the total energy of the system of two DOFs, as the participation number $N$ is held constant. The normalization is needed since the total energy is not constant when searching for the initial condition while holding $N$ fixed. In the next section, the values of $\xi_{\text{RM}}$ and $\xi^{\text{inf}}_{\text{IC}}$ are calculated for the limit of vanishing foundation (proving to be close), the correct value of $\hat{J}_1$ for the initial condition of the LPT is obtained and the critical coupling value is derived.

\section{The limit of vanishing foundation -- initial conditions for the (generalized) LPT and derivation of the critical coupling}
\label{Sec6}

\subsection{Initial conditions for the (generalized) LPT}

For the limit of vanishing foundation, the kinetic energy of the second oscillator in the case of localization is much larger then the rest of the energy, and thus its maximization is almost insensitive to $\gamma$. Therefore it is beneficial to follow the formally equivalent rout of the minimization of the rest of the energy, namely the potential energy of the second oscillator, plus the energy of the first oscillator and the coupling energy. The potential energy of the second oscillator and the coupling energy both depend on the average of the square of the second coordinate, $\langle\hat{q}_2^2\rangle_{_{\Theta}}$. One would expect this value to be $1/3$, since it is the average of a square of what in the case of vanishing foundation is a linear function (on correctly chosen single half-period). Indeed, direct calculation of the limit of Eqs. (\ref{eq22}), (\ref{eq33}) and (\ref{eq34}) shows that in the vanishing foundation limit one obtains:
\begin{equation}
\begin{split}
\label{eq35}
\nu_2(\gamma_0) \triangleq \frac{2}{\pi}\arcsin\frac{1}{\sqrt{\max\left(1,\hat{E}_{2,0}\right)}}\underset{\hat{E}_{2,0}\gg 1} \to 0 \\
\langle\hat{q}_2^2\rangle_{_{\Theta}}=\frac{8}{\pi^2} \sum_{n=1,3,5}^{\infty}\frac{\nu_2^2(\gamma)\cot^2{\frac{\pi\nu_2(\gamma)}{2}}}{\left[n^2-\nu_2^2(\gamma)\right]^2}\underset{\nu_2(\gamma_0)\to 0}\to
\frac{8}{\pi^2} \frac{4}{\pi^2}\sum_{n=1,3,5}^{\infty}\frac{1}{n^4}=\frac{1}{3}
\end{split}
\end{equation}

Consequently, one obtains the result $\xi^{\text{inf}}_{\text{IC}}=\sqrt{2/3}\approx 0.8165$.

The energy of the first oscillator in the case of localization is simply $\hat{E}_{1,0}=\hat{J}_{1,0}$ (no impacts). As established earlier, one can write $\langle\hat{q}_1^2\rangle_{_{\Theta}}=\frac{1}{2}\hat{J}_{1,0}$. The quantity left to be calculated is the coupling term, namely, $\langle\hat{q}_1\hat{q}_2\rangle_{_{\Theta}}$.
As $\nu_{2,0}\underset{\hat{E}_{2,0}\gg 1} \to 0,\nu_{1,0}\underset{\hat{E}_{1,0}< 1} = 1$, one has from Eq. (\ref{eq34}) that
\begin{equation}
\begin{split}
\label{eq36}
\langle\hat{q}_1\hat{q}_2\rangle_{_{\Theta}}=\frac{8}{\pi^2}\sqrt{\hat E_{1,0}}  \sum_{n=1,3,5}^{\infty}\frac{\nu_{1,0}\cos{\frac{\pi\nu_{1,0}}{2}}}{n^2-\nu_{1,0}^2}\frac{\nu_{2,0}\cot{\frac{\pi\nu_{2,0}}{2}}}{n^2-\nu_{2,0}^2} \underset{\nu_{2,0}\to 0,\nu_{1,0}\to1}\to\frac{4}{\pi^2}\sqrt{\hat J_{1,0}}
\end{split}
\end{equation}

Consequently, one obtains the result $\xi_{\text{RM}}=8/\pi^2\approx 0.8106$. Therefore we have established that the unknown coefficient in the initial Hamiltonian should be in the extremely narrow range (relatively to other variables in the brackets corresponding to the coupling energy, which are of the order of unity) of (0.8106,0.8165]. In such a narrow range, with a smooth dependence on $\hat{J}_1$ far enough from 0, it seems justified to pick any value, including that of the lower bound, which is what was done and proved to produce good results.

Henceforth, the energy complementary to the kinetic energy of the second oscillator is expressed as follows:
\begin{equation}
\begin{split}
\label{eq37}
\hat{E}_c\underset{k_1\to 0}\to\frac{1}{3}+\hat{J}_{1,0}+\hat{k}\left( \frac{1}{3}+\frac{1}{2}\hat{J}_{1,0}- \frac{8}{\pi^2}{\hat J_{1,0}}^{\frac{1}{2}}\right)
\end{split}
\end{equation}

For numerical maximization, it is beneficial that the maximized quantity stays of order unity, and hence the normalized form given in Eq. (\ref{eq34}). However, for analytic maximization, the normalization is unnecessary as long as the total energy is assumed constant. In the considered asymptotic limit, the remaining kinetic energy of the second oscillator is asymptotically equal to the total energy and remains constant regardless of the value of $\hat{J}_{1,0}$, as long as it is below unity. Thus the non-normalized energy in Eq. (\ref{eq37}) is sufficient.

Moreover, it is evident that the critical coupling in the considered system should be such that the coupling energy would be of the order of the kinetic energy of the second oscillator. Therefore, when normalized by the foundation stiffness $k_1$, the critical coupling should satisfy $\hat{k}_{\text{cr}}={k}_{\text{cr}}/k_1 \underset{k_1\to 0}\gg 1$. Furthermore,  in the case of only one particle experiencing impacts, the displacements of the two particles cannot be (even nearly) identical, and therefore the term in the parentheses in Eq. (\ref{eq37} should be of finite order. This means that in the considered limit in the critical case the coupling energy is the dominant part of the complementary energy and the first two terms in  Eq. (\ref{eq37} are negligible.

Therefore, the function to be minimized (corresponding to minimizing initial delocalization) for obtaining correct initial conditions for a generalized LPT is the term in the parentheses in  Eq. (\ref{eq37}:
\begin{equation}
\begin{split}
\label{eq38}
\hat{E}_M(\hat{J}_{1,0})\underset{k_1\to 0}\to\frac{1}{3}+\frac{1}{2}\hat{J}_{1,0}-\frac{8}{\pi^2}\sqrt{\hat J_{1,0}}
\end{split}
\end{equation}
where there is the constraint $\hat{J}_{1,0}\in[0,1]$.
The minimum point of the function in Eq. (\ref{eq38}) is
\begin{equation}
\label{eq39}
\begin{split}
\hat J_{1,0}^{\hat{E}_M^{min}}\underset{k_1\to 0}\to \underset{\hat{J}_{1,0}\in[0,1]}{\text{argmin}}\left(\frac{1}{2}\hat{J}_{1,0}-\frac{8}{\pi^2}\sqrt{\hat J_{1,0}}\right) =\frac{64}{\pi^4} \approx 0.657
\end{split}
\end{equation}

The corresponding finite though small minimum delocalization (coupling) energy (in $\hat{k}$ units) reads:
\begin{equation}
\label{eq40}
\begin{split}
\hat{E}_M^{min}\underset{k_1\to 0}\to \frac{1}{3}-\frac{32}{\pi^4}\approx  0.0048
\end{split}
\end{equation}

Had we used $\xi^{\text{inf}}_{\text{IC}}$ instead of $\xi_{\text{RM}}$, we would have gotten the result $\hat J_{1,0}^{\hat{E}_M^{min}}\underset{k_1\to 0}\to2/3\approx 0.667$ instead of 0.657.

Next, employing Eqs. (\ref{eq21}--\ref{eq23}), (\ref{eq30}), (\ref{eq37}), (\ref{eq39}) and (\ref{eq40}), we obtain the initial condition for the averaged Hamiltonian corresponding to the (generalized) LPT:
\begin{equation}
\begin{split}
\label{eq41}
h_{\text{cr}}(\gamma_0\to 0,\vartheta=0)\underset{k_1\to 0}\to\frac{64}{\pi^4}+\frac{\pi^2}{16}\left(N^2-\frac{64}{\pi^4}\right)^2+\frac{1}{3}
+\frac{16}{45\pi^2}\left(N^2-\frac{64}{\pi^4}\right)^{-2}+\\+\left( \frac{1}{3}-\frac{32}{\pi^4}\right)\hat{k}_{\text{cr}} \underset{N\gg 1}\to  \frac{\pi^2}{16}N^4+\left( \frac{1}{3}-\frac{32}{\pi^4}\right)\hat{k}_{\text{cr}}
\end{split}
\end{equation}

The notation $h_{\text{cr}}(0,0)$ was used in Eq. (\ref{eq41}) due to the fact that $\sin^2{\frac{\gamma_0}{2}}=\hat{J}_{1,0}/N^2=\frac{64}{\pi^4N^2}\underset{N\gg 1}\to 0$ and hence $\gamma
_0\to0$ (and the choice $\vartheta_0=0$ was already justified earlier). The value of $N$ becomes very large for vanishing foundation stiffness since $N^2=\hat{J}_1+\hat{J}_2$ and the actions are both normalized by $k_1$ and thus their sum becomes large for $k_1\to0$ since in the impact-regime at least one of the (non-normalized) actions is at least finite (and positive).

The subscript `$\text{cr}$' was used for both the Hamiltonian and the coupling due to the fact that some of the assumptions used in the asymptotic derivations presented above are justifiable only in the case of considerable energy exchange between the modes, which for the generalized LPT corresponds to the critical case.

The fact that the result $\gamma_0\to0$ was obtained means, in a sense, that the LPT in the considered system is similar to what was studied in other systems, including in \cite{Sapsis2017}. However in the case considered here, the condition $\gamma_0\to0$ does not correspond to action localization in the sense that $J_1=0$, which was the case in classic applications of the approach. In this sense the definition presented in this and the previous sections is a generalization of the concept of LPT.

In conclusion of this subsection, it would be instrumental to relate $N$ to the initial conditions in the original system. If we wish the LPT or the generalized LPT to represent the impulsive perfectly localized initial conditions in the \emph{full} system, then we should assume that initially the energy of the full system is only kinetic and is contained in the second oscillator, and can be expressed as: $E_0=\frac{1}{2}mV_0^2$, with $V_0$ being the velocity initially given to the second oscillator, with the first oscillator initially at rest and all three springs initially relaxed. Then, in normalized form one would have: $\hat{E}_0=\frac{mV_0^2}{k_1d^2}$. Since the averaging cannot change the energy, $h_{\text{cr}}(0,0)=\hat{E}_0$ would hold. Therefore one obtains:
\begin{equation}
\label{eq42}
\begin{split}
 \frac{\pi^2}{16}N^4 =\frac{mV_0^2}{k_1d^2}-\left( \frac{1}{3}-\frac{32}{\pi^4}\right)\hat{k}_{\text{cr}}
\end{split}
\end{equation}

It is noteworthy that the perfectly localized impulsive initial conditions naturally feasible for the full system are not in fact represented in the averaged system (the reason is basically that the squared sine has a nonzero average on a period). This fact poses no problem for the identification of the LPT for many systems, among others the case studied in \cite{Sapsis2017}, since the nonzero addition to the energy augmenting the localized-impulsive one is independent of the averaged-system's state variables. Here, however, the fact that localized impulsive initial conditions are not represented in the averaged system does raise a question on correct identification of the LPT. The answer to this question is suggested in Eq. (\ref{eq34}).

\subsection{Identification of the saddle point and derivation of the critical coupling}

One observes from Eq. (\ref{eq23}) that there is a fixed-point in the averaged Hamiltonian corresponding to $\vartheta=\pi$. Moreover, the averaged Hamiltonian is clearly symmetric with respect to a switch between the oscillators, and thus a line in the middle of the range representing the distribution of energy between the oscillators is a symmetry plane, and this line corresponds to $\gamma=\pi/2$, which is thus where the associated derivative of the Hamiltonian vanishes. Therefore, a fixed point of the system, specified by $\partial{h}/\partial{\vartheta}=\partial{h}/\partial{\gamma}=0$, can be found at the point $(\gamma_*=\pi/2,\vartheta_*=\pi)$.

Furthermore, one observes that the Hamiltonian only depends on $\vartheta$ through the coupling energy, which is maximal when the particles are in the anti-phase, which occurs at $\vartheta=\pi$. Moreover, the individual energies of the oscillators at the symmetry plane are convex functions of the actions, and thus should have a minimum with respect to $\gamma$ at the fixed-point, at least for a large enough total energy. Therefore, for large enough energies there is a \emph{saddle} point at the value of $h(\gamma_*=\pi/2,\vartheta_*=\pi)$.

The saddle coupling energy now becomes: $\hat{k}\langle(\hat{q}_1+\hat{q}_2)^2\rangle_{_{\Theta}}=\hat{k}\langle(\hat{q}_1+\hat{q}_1)^2\rangle_{_{\Theta}}=4\hat{k}\langle\hat{q}_1^2\rangle_{_{\Theta}}$, which by employing Eq. (\ref{eq35}) is evident to be equal to $=4\hat{k}/3$. Next, it is clear that in the limit of vanishing foundation, the first term in the expansion in Eq. (\ref{eq30}) is sufficient to describe the energy of an individual oscillator. For symmetric action distribution one has: $\hat{J}_{1}=\hat{J}_{2}=N^2/2$ (where Eq. \ref{eq19}) is employed). Consequently, $\hat{J}_{1}^2=\hat{J}_{2}^2=N^4/4$ and thus one obtains the following result:
\begin{equation}
\label{eq43}
\begin{split}
h(\pi/2,\pi)\underset{k_1\to 0}\to\frac{\pi^2}{32}N^4+\frac{4}{3}\hat{k}
\end{split}
\end{equation}

By using similar steps, it is possible to obtain the averaged Hamiltonian in a neighborhood of the suspected saddle point, say in $\gamma\in[\pi/4,3\pi/4]$. The specification of the neighborhood for $\gamma$ (in essence, the requirement of being far enough from zero) is needed for the asymptotic limit of $N\gg1$ to yield $\hat{E}_{1,2}\gg1$ and subsequently the result $\nu_{1,2}\to 0$, which eliminates the $\gamma$-$\vartheta$ coupling term and yields the limit-form of the Hamiltonian. This functional form of the Hamiltonian is not particularly valuable so as to be presented here, since for numerical analysis the full form is more robust for correct representation in the entire range. However, this asymptotic form can be used to obtain the derivatives of the Hamiltonian at the suspected saddle, which read
\begin{equation}
\label{eq44}
\begin{split}
\left. \frac{\partial h}{\partial{\vartheta}}\right\rvert_{\vartheta=\pi}= \left. \frac{\partial h}{\partial{\gamma}}\right\rvert_{\gamma=\frac{\pi}{2}}= \frac{\partial ^2h}{\partial{\gamma}\partial\vartheta}\underset{k_1\to 0}\to 0 , \left. \frac{\partial ^2h}{\partial{\gamma}^2}\right\rvert_{\gamma=\frac{\pi}{2}}\underset{k_1\to 0}\to \frac{\pi^2}{16}N^4 , \ \left. \frac{\partial ^2h}{\partial{\vartheta}^2}\right\rvert_{\vartheta={\pi}}\underset{k_1\to 0}\to -\frac{8}{\pi^2}\hat{k}
\end{split}
\end{equation}

The derivatives in Eq. (\ref{eq44}) clearly indicate that the point $(\gamma_*=\pi/2,\vartheta_*=\pi)$ is a saddle point of $h(\gamma,\vartheta)$.

Next we assume the existence of the trajectory in the averaged flow going from the initial condition of the generalized LPT to the suspected saddle point, and equate the saddle point energy for the critical coupling case to the expression given in the end of Eq. (\ref{eq41}), the conservation of energy being necessary for autonomous delocalization-into-beating. The result reads
\begin{equation}
\begin{split}
\label{eq45}
\frac{\pi^2}{16}N^4+\left( \frac{1}{3}-\frac{32}{\pi^4}\right)\hat{k}_{\text{cr}} =\frac{\pi^2}{32}N^4+\frac{4}{3}\hat{k}_{\text{cr}} \\
\end{split}
\end{equation}

This relation, combined with Eq. (\ref{eq42}) produces an expression for the (not normalized) critical coupling for beating/delocalization in the limit of vanishing foundation:
\begin{equation}
\label{eq46}
{k}_{\text{cr}} \underset{k_1\to 0}\to\frac{3\pi^4}{96+7\pi^4}\frac{mV_0^2}{d^2} \approx 0.3757\frac{mV_0^2}{d^2}
\end{equation}
(one recalls the definition $\hat{k}\triangleq k/k_1$). In \cite{Sapsis2017}, a similar problem was solved while explicitly removing the foundation from the formulation. The result obtained therein was: ${k}_{\text{cr}} \underset{k_1\equiv 0}= \frac{3}{8}\frac{mV_0^2}{d^2}$ (keeping the notation and dimensions used in the present paper). One notes that the prefactor $3/8=0.375$ is very close to the value of 0.3757 obtained herein. Had we used $\xi^{\text{inf}}_{\text{IC}}$ instead of $\xi_{\text{RM}}$ for the initialization of the LPT, we would have obtained exactly that result of a prefactor of $3/8$. This means that in \cite{Sapsis2017}, in essence the infimum for the Hamiltonian was used for initialization of initial conditions, instead of the RM Hamiltonian. It was possible there not to remain on the RM for the derivation of the critical coupling due to the fact that the critical coupling \emph{prefactor}there did not depend on $N$. Thus, there is no jump at the limit of vanishing foundation, but rather a different choice of the initial point dictated by the dependence on $N$. However, had we not used a generalization of the underlying conditions of an LPT, and simply taken $\hat{J}_{1,0}=0$, by analogy with what was done in \cite{Sapsis2017}, the numeric prefactor for the critical coupling would have been $3/7\approx 0.4286$, which is relatively quite far both from the value of $3/8$ of the case of no foundation per se and the result given in Eq. (\ref{eq46}).

\begin{figure}[!h]
\begin{center}
{{\includegraphics[scale = 0.33,trim={0.5cm 0 1.32cm 2},clip]{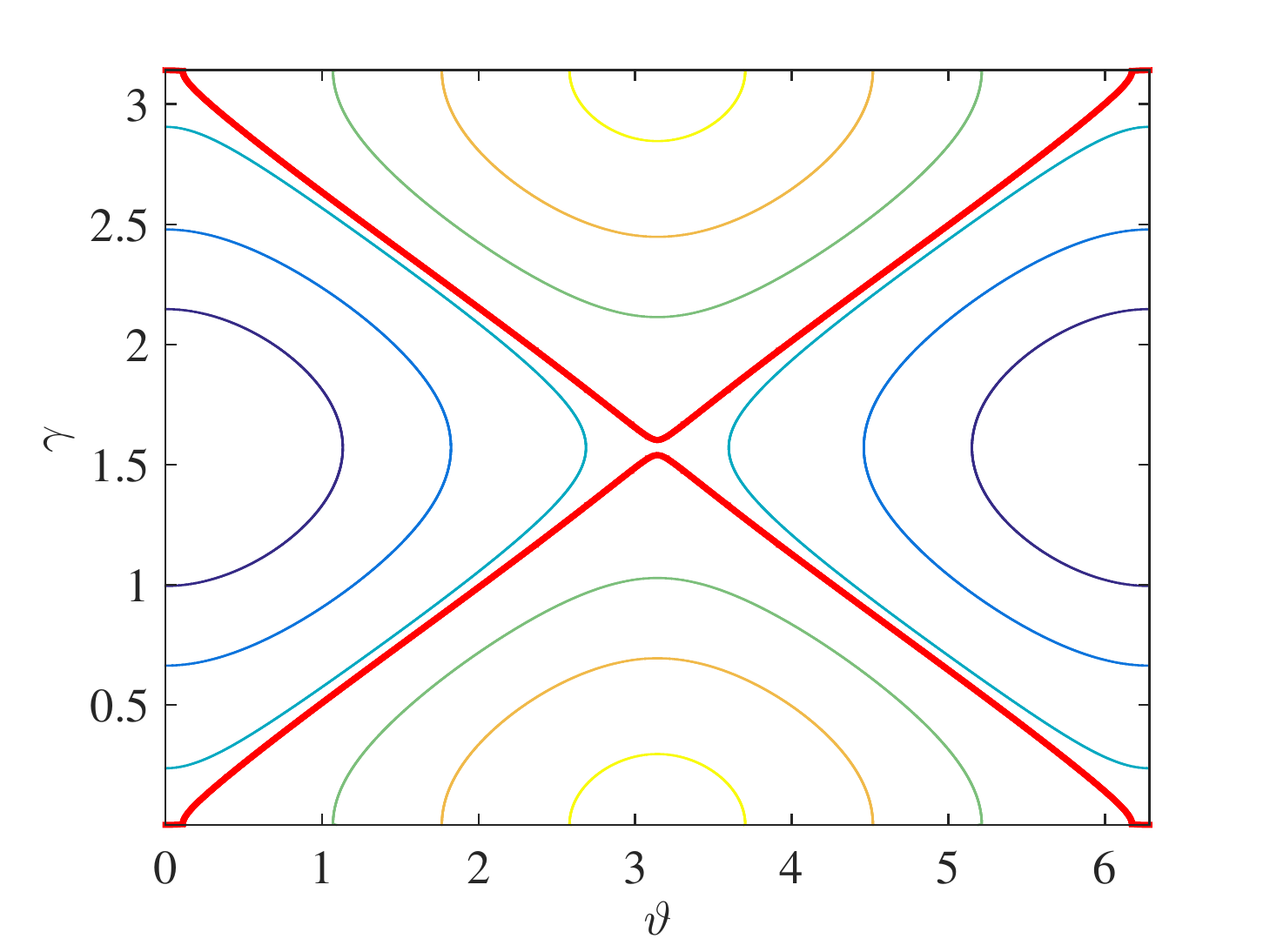}}
{\includegraphics[scale = 0.33,trim={0.5cm 0 1.32cm 2},clip]{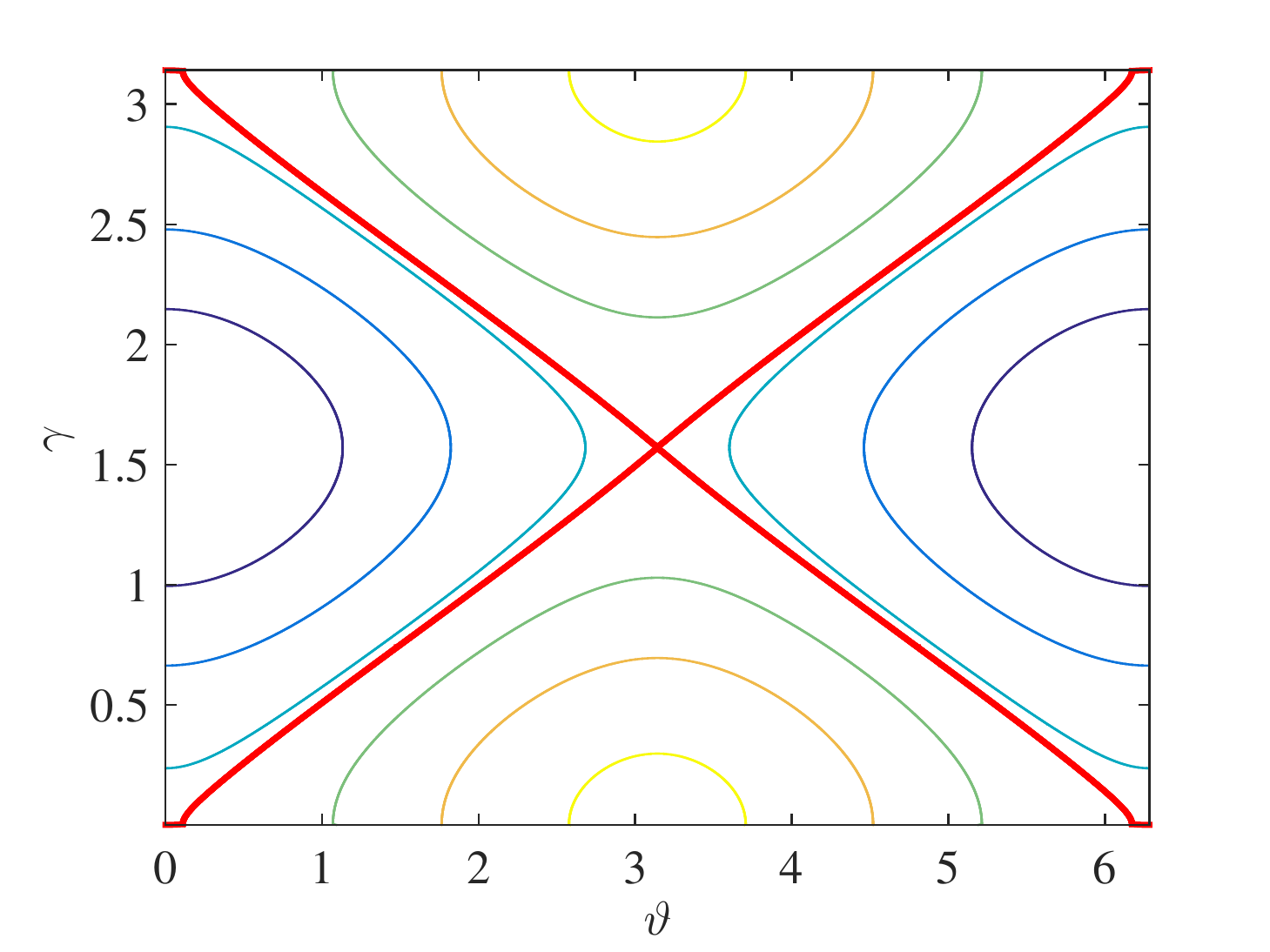}}
{\includegraphics[scale = 0.33,trim={0.5cm 0 1.32cm 2},clip]{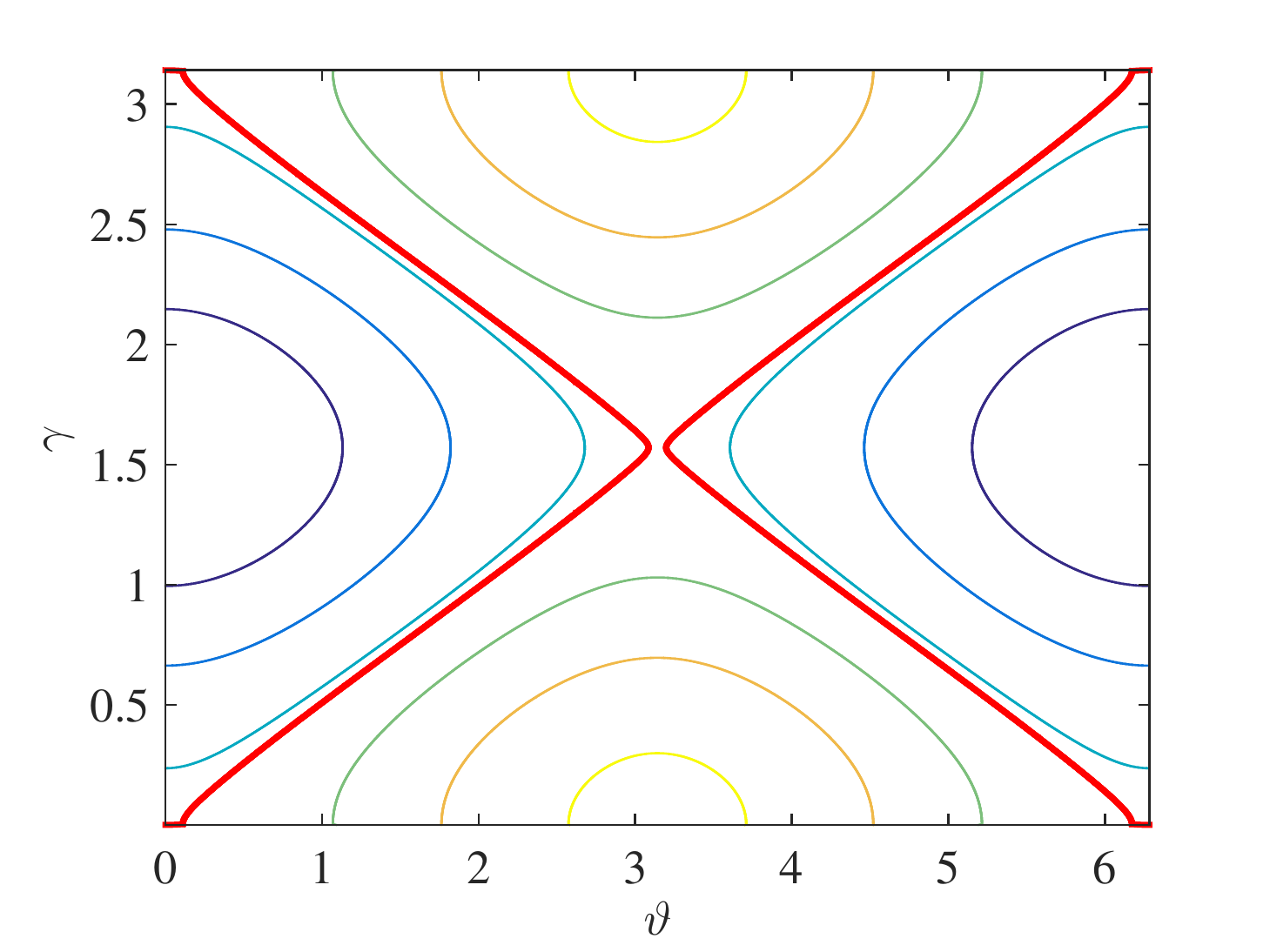}}}
\end{center}
\caption{\small Contour lines of $h(\gamma,\vartheta)$, with (generalized) LPT in thick (red online), for $\hat{k}=0.999\hat{k}_{\text{cr}}$ (left), $\hat{k}=\hat{k}_{\text{cr}}$ (middle), and $\hat{k}=1.001\hat{k}_{\text{cr}}$ (right), where $\hat{k}_{\text{cr}}$ is given in Eq. (\ref{eq46}), and $N=10^4$}
\label{Fig5}
\end{figure}

Illustration of the coincidence of the generalized LPT with the saddle point for the critical coupling value in the vanishing foundation limit is shown in Fig. \ref{Fig5}.

For the subcritical coupling, a system starting at utmost-localization conditions reaches a point of equal distribution of action in the anti-phase, and then goes back to localization at the same particle as in the beginning, completing a $2\pi$ phase-difference travel. During this process, marked by the thick (red online) curve, the other particle never attains a majority of the overall energy. In contrast to that, for slightly supercritical coupling, the phase never supersedes $\pi$, whereas $\gamma$ travels through its entire range and energy flows from one particle, which has a majority of the energy in the beginning, to the other particle, which attains a majority of it in the end of the period (with corresponding active impact-dynamics).

One notes the initial part of the orbit, corresponding to $\gamma \ll1$, which is flat, much unlike the majority of the curve, which is nearly straight with a slope of $1/2$ for about half of its range. This illustrates the fact that the asymptotic form of the Hamiltonian, with the vanishing $\gamma$-$\vartheta$ coupling, is incorrect for $\gamma\ll 1$. One other important point to be made here concerns the limit of \emph{vanishing impact velocity}. Clearly, this is an opposite limit to what was examined previously in this section, namely $\hat{E}\to 1^+$ instead of $\hat{E}\gg 1$. One should expect that in this limit, where the nonlinear effects are vanishing, there would be no localization regime at all due to equipartition of energy. In other words, one should expect delocalization here for a coupling value as small as zero. Therefore, looking at the entire range, the critical coupling $\hat{k}_{\text{cr}}$ as a function of the total normalized energy $\hat{E}$, should be a function starting at zero for $\hat{E}=1$ and increasing until reaching a horizontal asymptote for $\hat{E}\gg 1$, with the value given in Eq. (\ref{eq46}).

The following section generalizes the calculation of the critical coupling for finite values of the foundation stiffness. It is subsequently validated that indeed the proposed generalization of the definition of the LPT allows good prediction of the critical transition of the unaveraged system.

\section{Critical coupling for finite foundation stiffness values -- numerical analysis}
\label{Sec7}

In this section the averaged system assumed close to 1:1 resonance is analyzed numerically, using the results obtained in the previous sections. The analysis is encompassed in obtaining the critical coupling stiffness, which corresponds to the value for which the saddle point of the averaged Hamiltonian coincides with the generalized LPT, starting from the point of maximum localization of energy in one particle's momentum. Clearly, for $\hat{E}=1$, the system is linear and there is no localization. For $\hat{E}\gg1$, which numerically occurs say for $\hat{E}>100$, the asymptotic result presented in the previous section is valid. Therefore an appropriate order of magnitude of the normalized energy for numerical analysis should be between 1 and 100, or in other words of the order of magnitude of $\hat{E}=10$. A reasonable range would be $\Delta\hat{E}\approx 10$, so as to stay far enough from both 1 and 100. In Fig. \ref{Fig6} below, the critical coupling, in units of $mV_0^2/d^2$ is given as function of $\hat{E}$ in the range of approximately $[6,16]$. The associated $\xi$-range for these energies, referring to Eq. (\ref{eqIC6}), never leaves the bounds of the interval [0.81,0.83], making the investigation on the RM justified for the finite foundation values calculations as well. The critical coupling values for transition from localization to delocalization are computed for 12 points, more or less uniformly spanning the range. The critical coupling values for the averaged system are given on top of the critical values calculated for the full system, where the criterion for delocalization was defined as the occurrence of impacts for both oscillators (this is further elaborated on in the next subsection).
\begin{figure}[!h]
\begin{center}
{{\includegraphics[scale = 0.5,trim={0 0 0 0},clip]{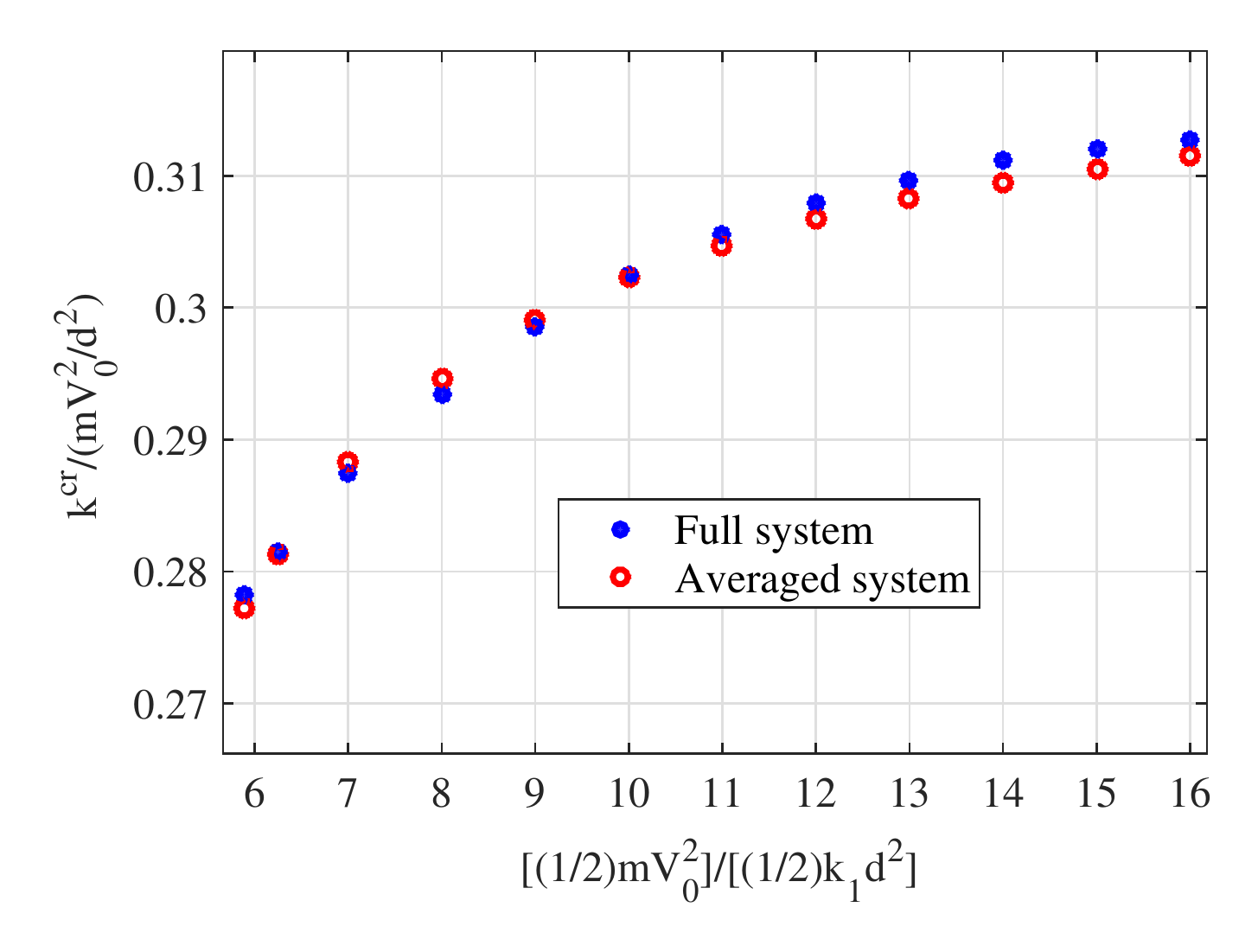}}}
\end{center}
\caption{\small Critical-coupling values for different energies. Full (blue online) circles represent the full system, and empty (red online) circles represent the averaged system with assumed 1:1 resonance}
\label{Fig6}
\end{figure}

Bearing in mind that in the limit $\hat{E}\to 1^+$, one has $k_{\text{cr}}\to 0$, and for $\hat{E}\to\infty$, one has $k_{\text{cr}} \approx0.3757mV_0^2/d^2$, one notes that Fig. \ref{Fig6} shows very good prediction of the critical transition of delocalization of the full system by the averaging approach combined with the notion of the generalized LPT, at least in the examined range. It is argued in the following that the chosen range is indeed a reasonable one.

The range to the left of what is shown in Fig. \ref{Fig6} is problematic, since close enough to the limit of vanishing impact velocity, the saddle point of the averaged Hamiltonian becomes very flat, on the verge of turning into a local maximum. This renders the corresponding generalized LPT extremely flat and the corresponding energy exchange - extremely slow. Consequently, one has to wait extremely long times for delocalization to occur, and therefore only for \emph{higher} coupling values one immediately observes the delocalization. Therefore, if judging by reasonable computational time length results, it may appear as if the averaging-based approach underestimates the critical coupling values.
The range to the right of what is shown in Fig. \ref{Fig6} is problematic for another reason. For high-enough energy it is suggested by numerical analysis that the full system undergoes delocalization for \emph{lower} coupling values than predicted by the averaging-based approach. The reason for this may be attributed to the fact that for high energies the system is highly nonlinear and higher resonances than the 1:1 resonance assumed to occur may come into play. Moreover there may be resonance overlap. Clearly if this happens for higher coupling than predicted by averaging, it is irrelevant to the onset of delocalization. If, however, it occurs for lower coupling values than predicted by assuming 1:1 resonance, then delocalization in the full system would occur for lower coupling values then predicted.

In the case analyzed in \cite{Sapsis2017}, this latter phenomenon did not occur. The reason is that when the system is rigorously foundation-free, the energy can be renormalized, such that the value of the energy has no influence on the dynamics and there is no high-energy regime with its possible complex resonance overlap.

A detailed analysis of the critical transition both for the averaged and for the full system, for a typical point within the range shown in Fig. \ref{Fig6}, is given in the following subsections.

\subsection{Example of numerical capture of the transition -- averaged system}

For detailed description of the numerical capture of the critical coupling, we chose the point $\hat{E}=9$, lying well within the range of good correspondence between the averaging-based and full critical analyses.

For the normalized total energy of $\hat{E}=[(1/2)mV_0^2]/[(1/2)k_1d^2]=9$, the critical coupling as obtained for the averaged system was found to correspond approximately to $k_{\text{cr}}\approx 0.299mV_0^2/d^2$.

\begin{figure}[!h]
\begin{center}
{{\includegraphics[scale = 0.48,trim={0 0 0 0},clip]{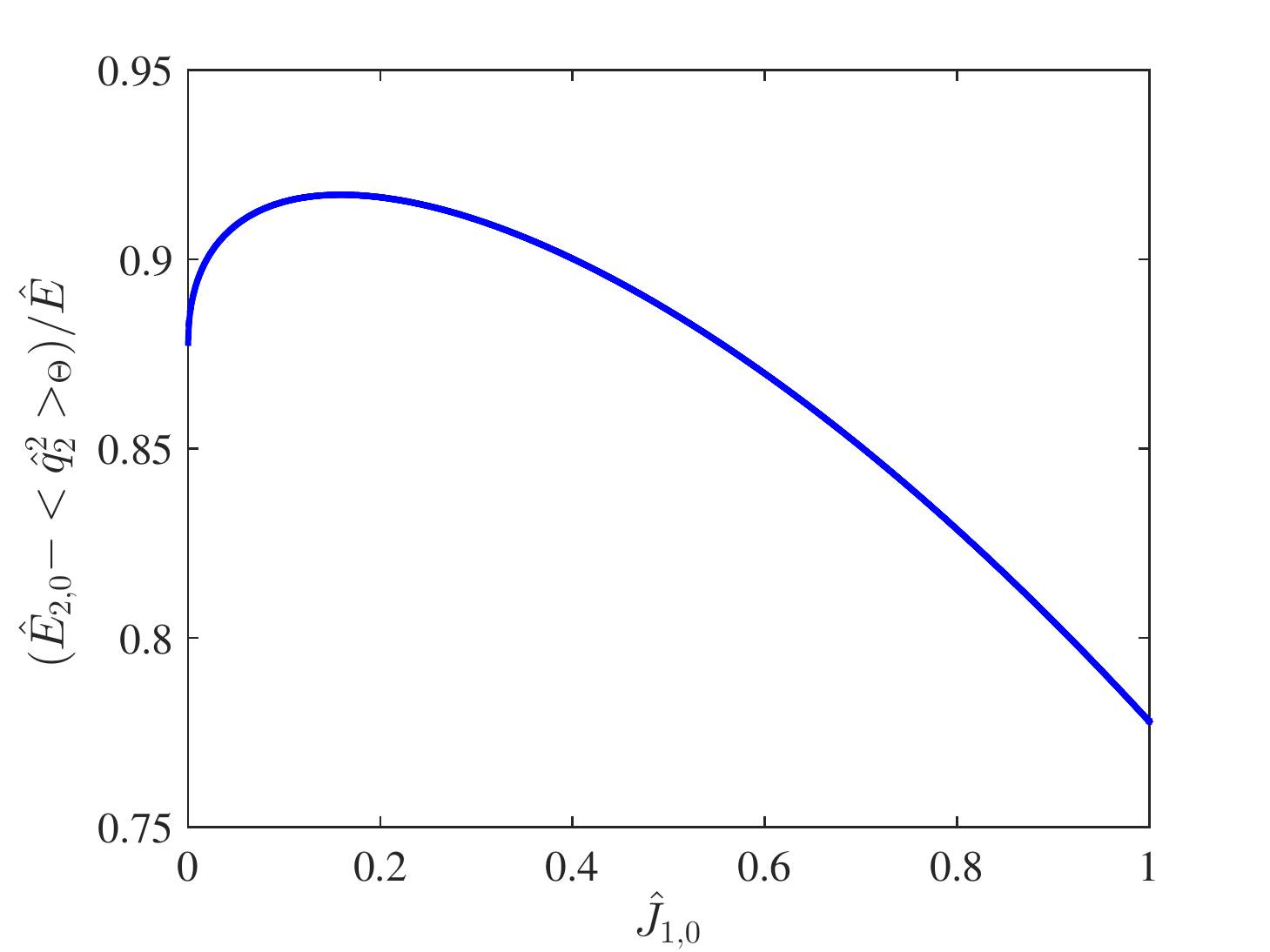}}}
\end{center}
\caption{\small Variation of normalized initial second-particle kinetic energy with $\hat{J}_{1,0}$, having a maximum at $\hat{J}_{1,0}\approx 0.1589$, plotted for $N \approx1.95,k_{cr}\approx0.29908 mV^2 0/d_2$ (the critical coupling for this energy).}
\label{Fig7}
\end{figure}

The first step in the process of determining this value was to calculate the correct $\gamma_0$, the argument parameterizing the initial first-particle action. The right  value of $\gamma_0$ was found by implementing Eq. \ref{eq34} by use of numerical line search. The objective function, namely the ratio of the initial kinetic energy of the second particle to the total energy for in-phase conditions as a function of $\hat{J}_{1,0}$ , in a range containing the maximum (for the correct value of the critical coupling and the associated value of $N$ to produce $\hat{E}=9$, found \emph{a posteriori}), is shown in Fig. \ref{Fig7}.

It is important to note that the maximum of the curve as shown in Fig. \ref{Fig7}, depends on the value of the coupling, which is not known in advance. Therefore, this is an iterative process. Moreover, the maximization is performed for a constant value of $N$. Thus this is a \emph{triple} loop iterative process. The outer loop is for the critical coupling. Inside this loop, for every iteration for the critical coupling there is a second loop for $N$. Inside the loop on $N$, for each value of $N$, there is a third loop, on $\gamma_0$. The loop on $\gamma_0$ is finished when for each choice of $N$ and $k$, the kinetic energy of the second oscillator normalized by the total energy is maximized. The middle loop on $N$ is terminated when a value of $N$ for each try of $k$ is found such that for $\gamma_{0,max}$ the total energy is equal to the assigned value of $\hat{E}$ (in our example it is 9). Therefore, in each iteration of the outer loop on $k$, one has $\gamma_0$ maximizing the ratio of the second-oscillator kinetic energy to the total energy, and $N$ making the corresponding total energy be of the correct, assigned value. The outer loop terminates when the generalized LPT obtained with the calculated $\gamma_0$ and $N$ passes through the saddle point of the Hamiltonian, which is calculated for every try of $k$ with the correct value of $N$, such that the total energy is as desired (9 in the example).  If the plotted generalized LPT appears to reach a maximum value of $\gamma$ at $\vartheta=\pi$ and returns to $\gamma_0$ for $\vartheta=2\pi$, then a higher coupling value is chosen and the process is repeated iteratively, until the generalized LPT curve reaches $\gamma=\pi-\gamma_0$.

The aforementioned triple-loop algorithm was executed for the case of $\hat{E}=9$ (and as demonstrated earlier in this section also for all the other values of the energy in the range shown in Fig. \ref{Fig6}). The obtained triplet-solution $\lbrace k_{\text{cr}},N_{\text{cr}},\gamma_{0,max}^{\text{cr}}\rbrace$ is $k_{\text{cr}}\approx 0.29908mV_0^2/d^2,N_{\text{cr}}\approx 1.953827,\gamma_{0,max}^{\text{cr}}\approx 0.4109$ ($\hat{J}_{1,0}\approx 0.1589$).

\begin{figure}[!h]
\begin{center}
{{\includegraphics[scale = 0.33,trim={0.5cm 0 1.32cm 2},clip]{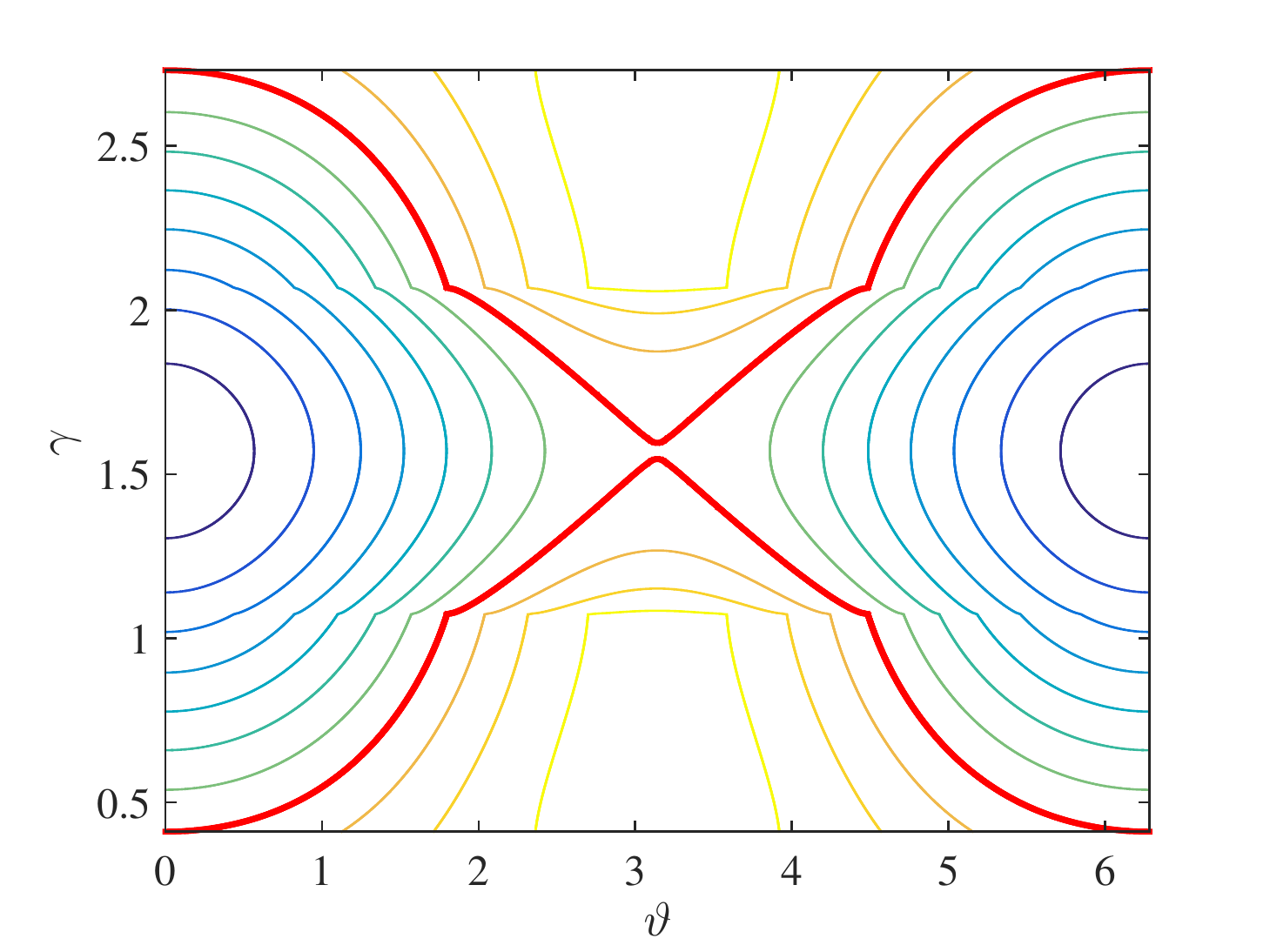}}
{\includegraphics[scale = 0.33,trim={0.5cm 0 1.32cm 2},clip]{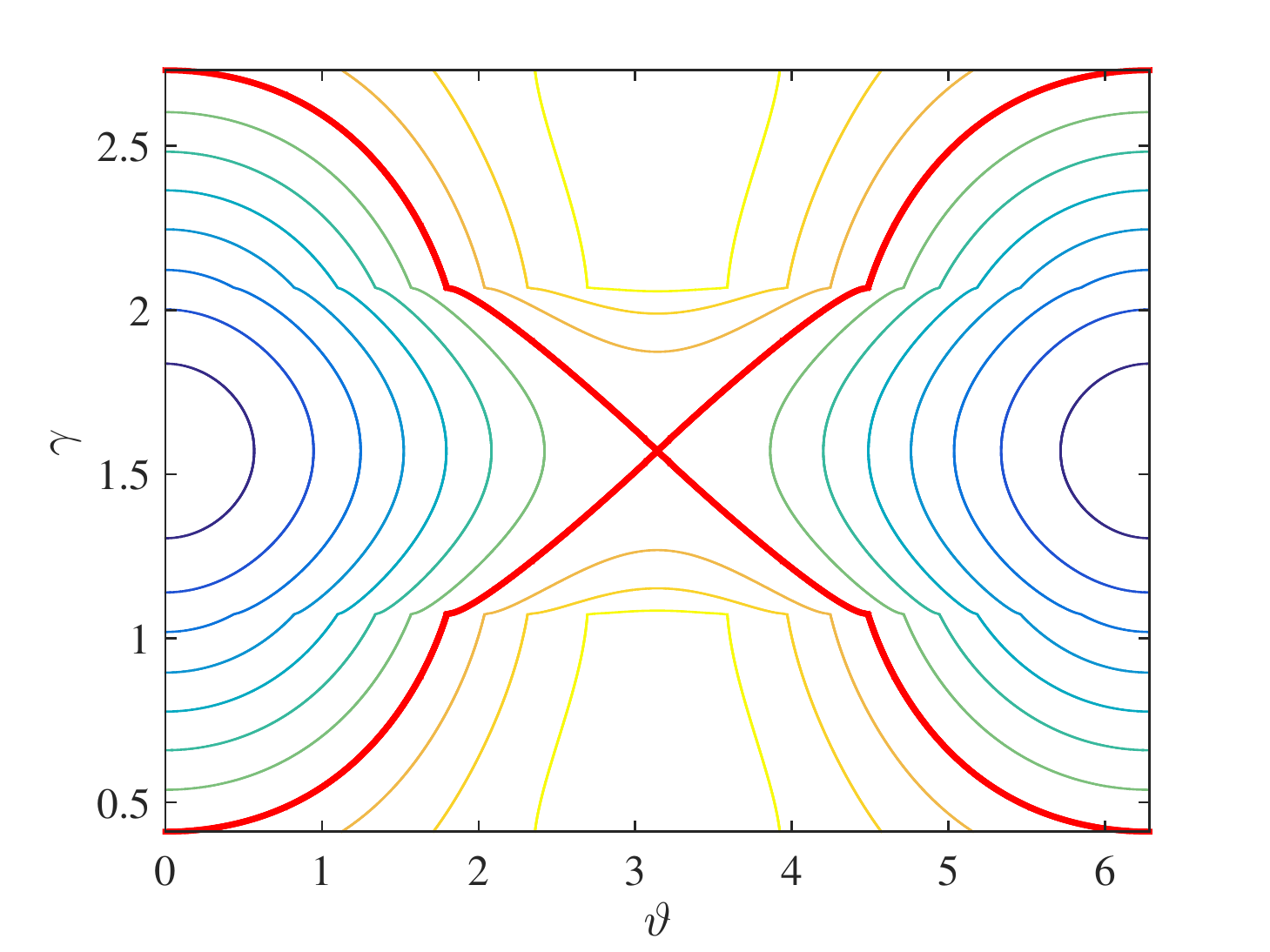}}
{\includegraphics[scale = 0.33,trim={0.5cm 0 1.32cm 2},clip]{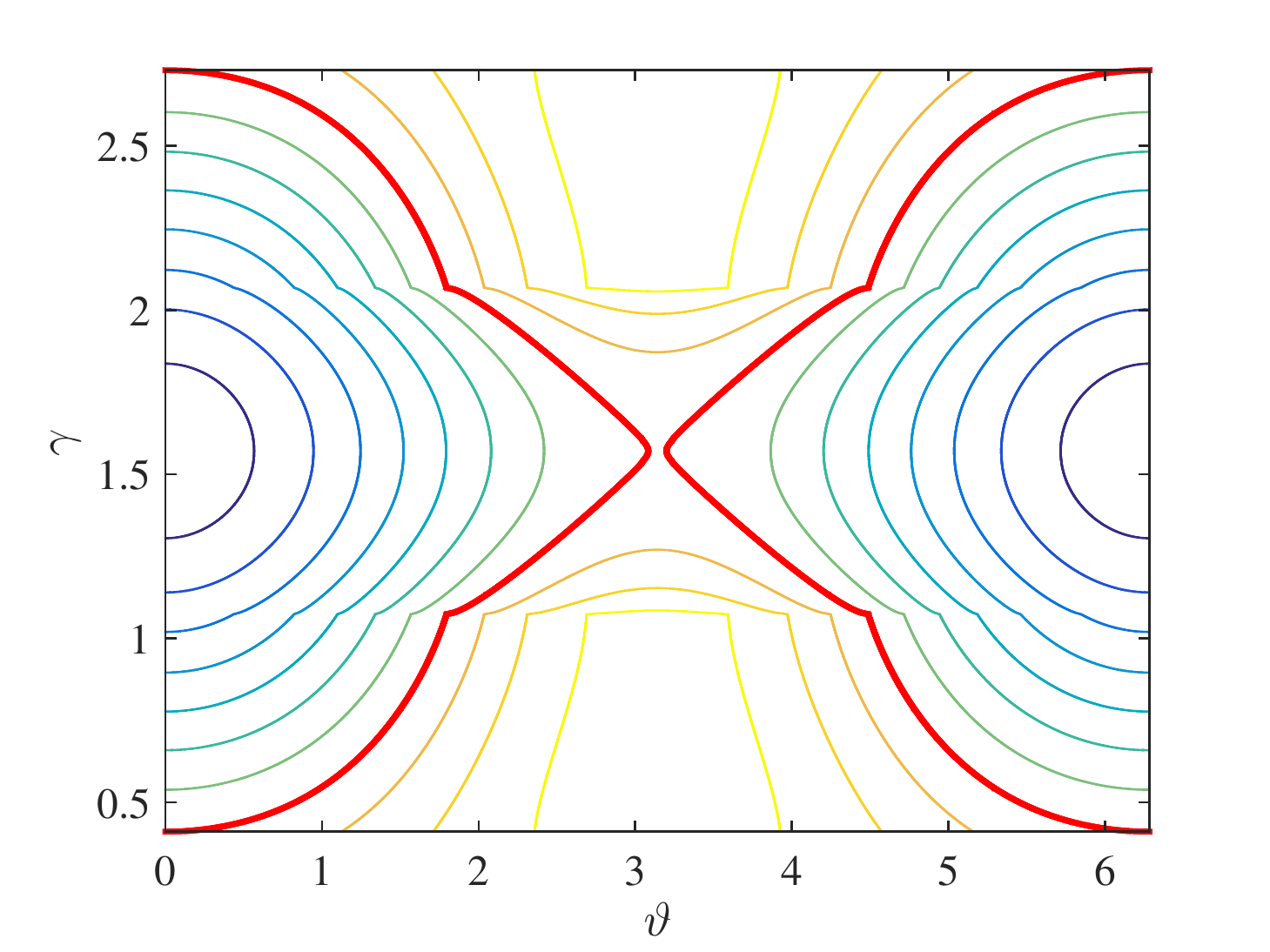}}}
\end{center}
\caption{\small Contour lines of $h(\gamma,\vartheta)$, with (generalized) LPT in thick (red online), for $\hat{k}=0.2988mV_0^2/d^2$ (left), $\hat{k}=\hat{k}_{\text{cr}}=0.29908mV_0^2/d^2$ (middle), and $\hat{k}=0.2994mV_0^2/d^2$ (right), in the case of $\hat{E}=9$}
\label{Fig8}
\end{figure}

Fig. \ref{Fig8} shows contour plots of the Hamiltonian, with around-critical generalized LPT curves for $\hat{E}=9$ (and the aforementioned corresponding parameter triplet value). One notes that for supercritical coupling the majority of the energy in the end of a slow-flow period is contained in the action of the first particle.

For comparison, had we not used the concept of the \emph{generalized} LPT, but rather employed the notion by na\"ively taking $\gamma_0=0$, we would have observed the critical transition of delocalization at the value $k_{\text{cr}}^{\gamma_0=0}\approx 0.349mV_0^2/d^2$, which is clearly much farther than the value $k_{\text{cr}}^{\text{g-LPT}}\approx0.29908mV_0^2/d^2$, obtained using the aforementioned maximization concept, from the value $k_{\text{cr}}^{\text{full}}\approx0.2985mV_0^2/d^2$, emerging for the full system (addressed below).

\subsection{Example of numerical capture of the transition -- full system}

\begin{figure}[!h]
\begin{center}
{{\includegraphics[scale = 0.45,trim={0 0 0 0},clip]{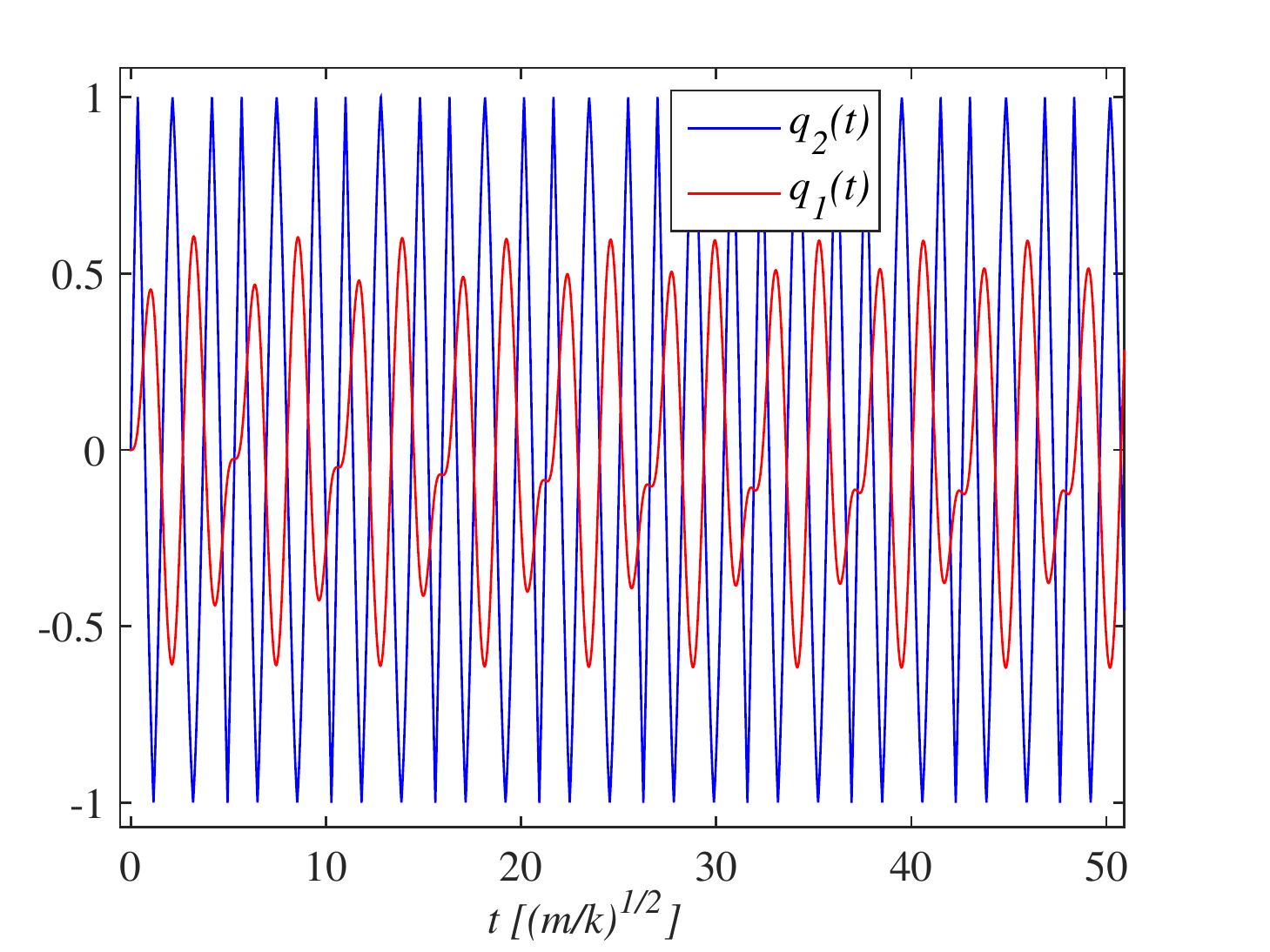}}
{\includegraphics[scale = 0.45,trim={0 0 0 0},clip]{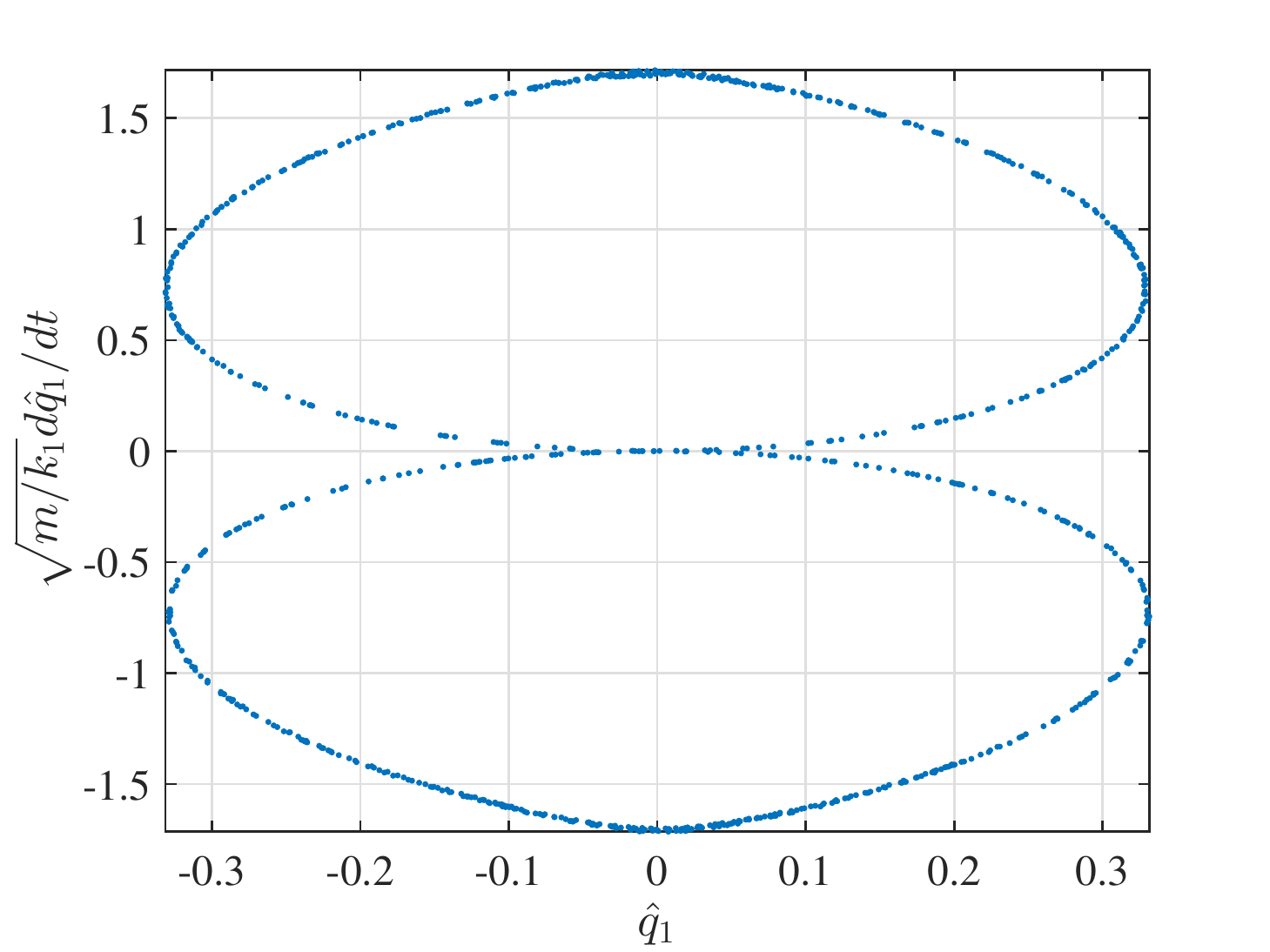}}}
\end{center}
\caption{\small Integration results for the full system with localized (on the second oscillator) impulsive initial conditions, for $\hat{E}=9,k_{\text{cr}}=0.298mV_0^2/d^2$,  showing displacements (left) and a Poincar\'e section defined by $q_2^{\text{PS}}=0$ (right) }
\label{Fig9}
\end{figure}

\begin{figure}[!h]
\begin{center}
{{\includegraphics[scale = 0.45,trim={0 0 0 0},clip]{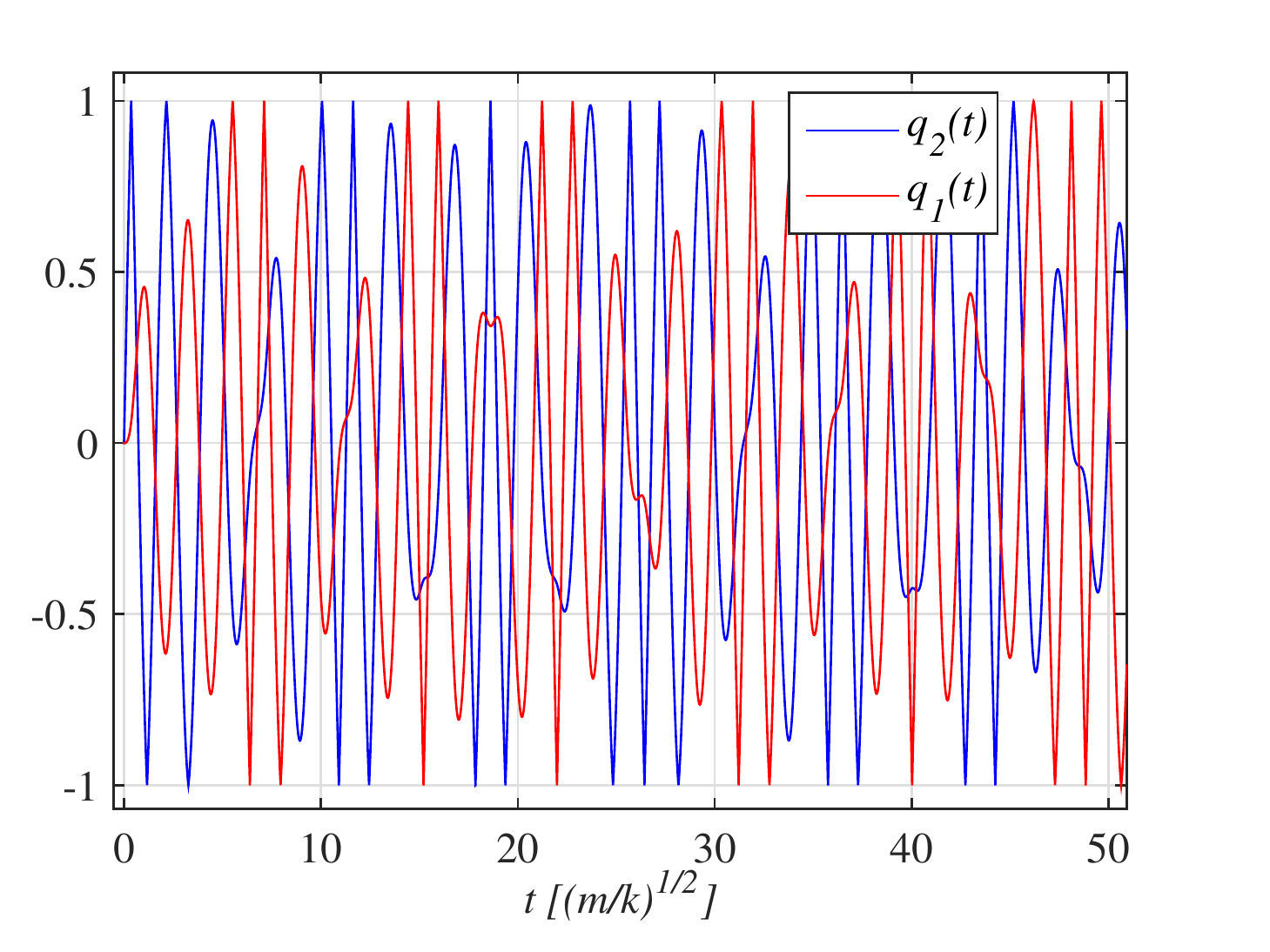}}
{\includegraphics[scale = 0.45,trim={0 0 0 0},clip]{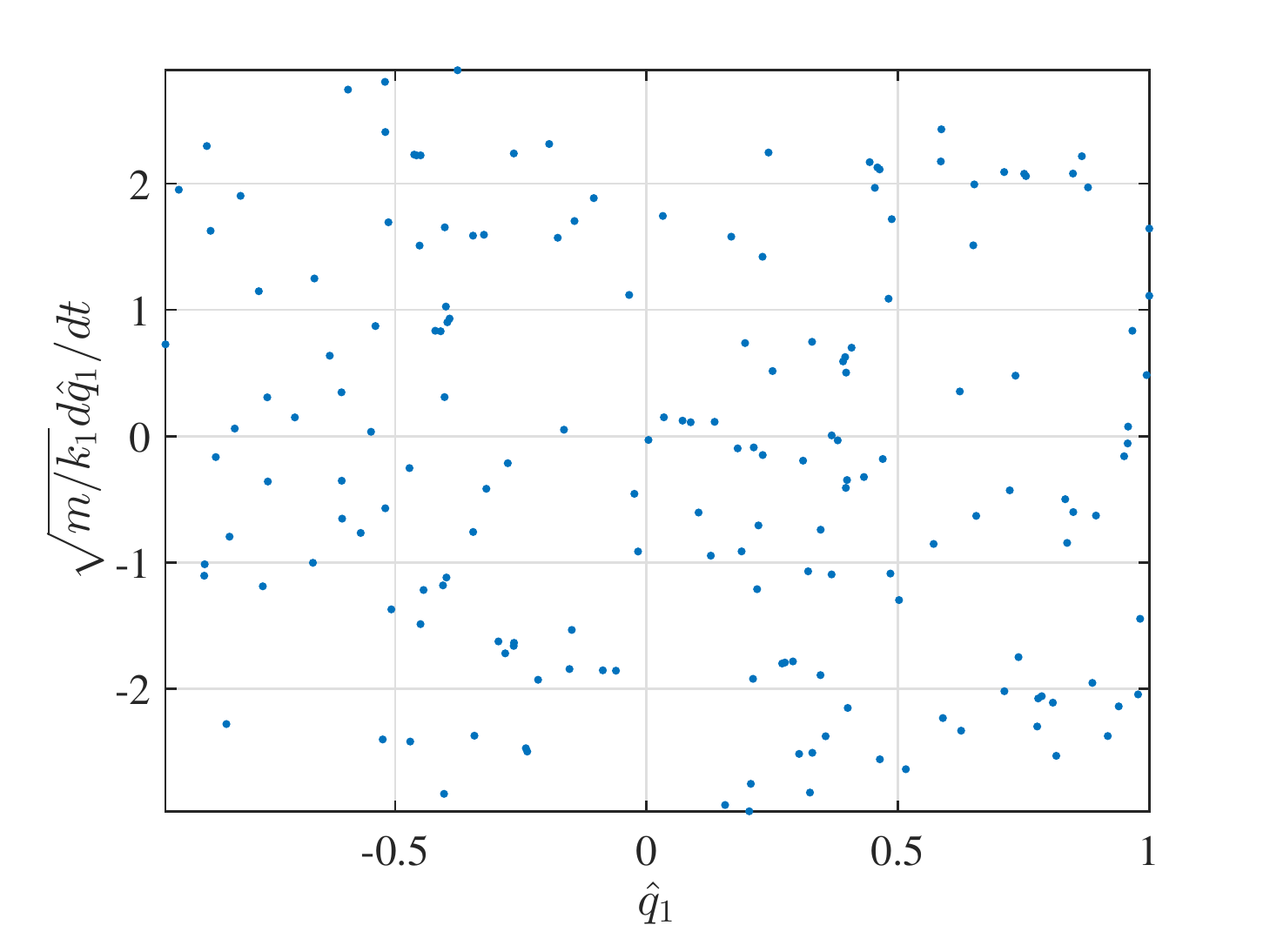}}}
\end{center}
\caption{\small Integration results for the full system with localized (on the second oscillator) impulsive initial conditions, for $\hat{E}=9,k_{\text{cr}}=0.299mV_0^2/d^2$,  showing displacements (left) and a Poincar\'e section defined by $q_2^{\text{PS}}=0$ (right) }
\label{Fig10}
\end{figure}

In order to estimate the success of capturing critical transitions by use of averaging with the generalized LPT concept suggested in this work, the full system was modeled numerically, with impacts being represented by a high odd-power displacement-dependent \emph{force} term appropriate for conservative impacts and suitable for a stable integration algorithm (here $-(2\xi+1)\hat q_{1,2}^{4\xi+1}$, with $\xi=500$). The solver that was employed for the purpose is a one-step method based on the trapezoidal rule and a backward differentiation formula of order 2 with a free interpolant \cite{Hosea1996}. The modeling was performed for various energies and the critical coupling values for delocalization were established and are shown in Fig. \ref{Fig6}. Detailed results for the case of $\hat{E}=9$ are shown in Figs. \ref{Fig9} and \ref{Fig10}.

Fig. \ref{Fig9} shows integration results for the full system of two dynamic degrees of freedom, which for constant energy occupies a three-dimensional manifold in the phase space. The left plot shows the displacement history for short times. The plot on the right shows a Poincar\'e section (PS) made at $q_2=0$. On the PS, the flow occupies a two-dimensional manifold and thus for long enough times the entire dynamics should be revealed there. One observes the two ovals typical for regular periodic dynamics. Since the coupling is on the verge of creating resonance, one observes some thickening of the periodic orbit at the top and at the bottom, which may be indicative of chaos on the separatrix. In this case, since the energy is large enough and the nonlinearity is that of impact, namely the strongest one possible, secondary resonances, although not yet significantly shifting the critical coupling for delocalization, can nevertheless produce secondary bifurcations with coincidence of multiple flat local extrema. Thus there may exist a homoclinic orbit (turning into a heteroclinic orbit as coupling increases from subcritical to supercritical) passing through the primary saddle on the two-dimensional manifold of the PS, and it may have associated chaos on the separatrix. For smaller coupling, the PS shows well-defined strict periodic orbits with no thickening, and for $k=0.299mV_0^2/d^2$, there is clear delocalization, as shown in Fig. \ref{Fig10}.

The left plot in Fig. \ref{Fig10} shows the displacements, which reveal that both particles experience impact. Comparison of the left plots in Figs. \ref{Fig9} and \ref{Fig10} gives a typical picture of a critical transition, where a shift from $k=0.298mV_0^2/d^2$ to $k=0.299mV_0^2/d^2$, creates a change in $\hat{q}_1^{\text{max}}$ from 0.6 to 1. The right plot in Fig. \ref{Fig10} shows that a small increase in the coupling changes the dynamics on a PS from a periodic orbit (with perhaps some initial chaos on the separatrix) to completely intractable global chaos not localized on a small compact subspase of the manifold.

Figs. \ref{Fig9} and \ref{Fig10} are qualitatively similar to the results given in the same context in \cite{Sapsis2017}, albeit for finite foundation stiffness.

One observes the excellent correspondence between Fig. \ref{Fig8} and Figs. \ref{Fig9} and \ref{Fig10}, confirming the occurrence of a critical transition in the full system in accordance with the value predicted by the averaging approach combined with the concept of the generalized LPT.

\section{Conclusions}
\label{Sec8}
The present paper exemplified the benefit of the canonical transformation to the action-angle variables for the prediction of critical transitions in small non-integrable systems with internal symmetry, using a model system of two linearly-coupled oscillators in vibro-impact on-site potential with the addition of a harmonic part. In this sense it is an extension of the previous work \cite{Sapsis2017} for the case of finite additional harmonic on-site potential. This modification requires one to introduce the asymptotic technique for proper inversion of the energy - action relation. Similar problem of inversion exists in many model potentials, and the suggested approach can be seen as possible remedy to overcome this problem. Besides, it turns out that the notion of LPT introduced in \cite{Manevitch2014} and employed in \cite{Sapsis2017} is insufficient. For the case where the coupling energy is dependent on the actions of the individual oscillators, a more delicate definition of the LPT is required. Such generalization was given and examined asymptotically for the appropriate limit case. Furthermore, the generalization of the concept of LPT was examined numerically for finite values of the foundation stiffness, and in relatively wide  range, and good prediction of the critical transition by the average flow was demonstrated. The quality of this prediction is notably superior to the one obtainable with na\"ive application of the concept of the LPT without the suggested generalization.

This work in a sense complements the analysis made in \cite{NPOVG2017} for a system of two cross-linked chains in combined harmonic and vibro-impact on-site potential, where quasi-beating was observed numerically in the high-frequency limit. The present work shows by analysis of the averaged flow for the unit-cell of the aforementioned chain that beating indeed takes place for high-enough coupling.

The problem addressed in the paper may be seen as a particular case of a more general issue, relevant for all explorations of resonant phenomena in nonlinear systems. The dimensionality of the resonant manifold is lower than the dimensionality of the complete state space of the problem -- this simplification is exactly the reason why one cares about resonances. It means, however, that one cannot claim that the system is immediately captured into the resonance under any initial conditions. Therefore, the relationship between the real initial conditions of the system, and the initial point to be attributed to the resonance manifold, may be highly nontrivial. The method to overcome this difficulty suggested in the paper involves optimizing the initial conditions on the resonant manifold itself, and this can be applicable also for more general and complicated systems that exhibit strong resonant energy transfer.

\enlargethispage{20pt}
\section*{Authors' contributions}
NP performed the analysis, derivations and numeric simulations, and drafted the main part of the manuscript. OG conceived and supervised the study, and drafted parts of the manuscript.  All auhtors read and approved the manuscript.

\section*{Acknowledgments}
The authors are very grateful to the Israel Science Foundation (grant 1696/17) for financial support of this work.



\begin{thebibliography}{54}

\bibitem{NayfehMook} A.H.Nayfeh and D.T.Mook, Nonlinear Oscillations, Wiley, New York, 1979.
\bibitem{NayfehBala} A.H.Nayfeh and B.Balachandran, Applied Nonlinear Dynamics: Analytical, Computational and Experimental Methods, Wiley, 1995.
\bibitem{Arnold1989} V.I.Arnold, Mathematical Methods of Classical Mechanics, Springer, Berlin, 1989.
\bibitem{Sanders2007} J.A.Sanders, F.Verhulst and J.Murdock, Averaging Methods in Nonlinear Dynamical Systems, Springer, 2007
\bibitem{Awrejcewicz2012} J.Awrejcewicz, I.V.Andrianov and L.I.Manevitch, Asymptotic Approach in Nonlinear Dynamics: New Trends and Applications, Springer, 2012.
\bibitem{Fidlin2006} A.Fidlin, Nonlinear Oscillations in Mechanical Engineering. Springer, Berlin, 2006
\bibitem{Babitsky1978} V.I.Babitsky Theory of Vibro-Impact Systems and Applications, Springer, Berlin, 1998 (Revised translation from Russian, Nauka, Moscow, 1978)
\bibitem{Pilipchuk2010} V.N.Pilipchuk, Nonlinear Dynamics: Between Linear and Impact Limits. Springer, Berlin, 2010.
\bibitem{Babitsky2014} V.I.Babitsky and V.R.Hiwarkar, Modelling of structures with developing discontinuity, Journal of Sound and Vibration, 333, pp. 5917-5938, 2014.
\bibitem{Hiwarkar2012}  V.R.Hiwarkar, V.I.Babitsky and V.V.Silberschmidt, Crack as modulator, detector and amplifier in structural health monitoring,  Journal of Sound and Vibration, 331, pp. 3587-3598, 2012.
\bibitem{Andreaus2007} U. Andreaus, P.Casini and F.Vestroni, Non-linear dynamics of a cracked cantilever beam under harmonic excitation, International Journal of Non-Linear Mechanics, 42, 566-575, 2007.
\bibitem{Rosenberg1962} R.M.Rosenberg, The normal modes of nonlinear n-degree-of –freedom systems, ASME Journal of Applied Mechanics, 29, 7-14, 1962.
\bibitem{Rand1974} R.H.Rand, A direct method for nonlinear normal modes, International Journal of Non-Linear Mechanics, 9, 363-368, 1974.
\bibitem{Vakakis1996} A.F.Vakakis, L.I.Manevitch, Y.V.Mikhlin, V.N.Pilipchuk and A.A.Zevin, Normal Modes and Localization in Nonlinear Systems, Wiley, New York, 1996.
\bibitem{Ablowitz1976} M.J. Ablowitz and J.F. Ladik, Nonlinear differential-difference equations and Fourier analysis, Journal of Mathematical Physics, 17, 1011, 1976.
\bibitem{Ovchinnikov1999} A.A. Ovchinnikov and S. Flach,  Discrete Breathers in Systems with Homogeneous Potentials: Analytic Solutions, Physical Review Letters, 83, 248-251, 1999.
\bibitem{Gendelman2013} O.V.Gendelman, Exact Solutions for Discrete Breathers in Forced-Damped Chain, Physical Review E, 87, 062911, 1-11, 2013.
\bibitem{Peeters2009} M. Peeters, R. Viguié, G. Sérandour, G. Kerschen and J.-C. Golinval, Nonlinear normal modes, Part II: Toward a practical computation using numerical continuation techniques, Mechanical Systems and Signal Processing, 23, 195-216, 2009.
\bibitem{Mikhlin1995} Yu.V. Mikhlin, Matching of local expansions in the theory of non-linear vibrations, Journal of Sound and Vibration, 182, 577-588, 1995.
\bibitem{Sire2005} Y. Sire and G.James, Numerical computation of travelling breathers in Klein–Gordon chains, Physica D, 204, 15-40, 2005.
\bibitem{Birkhoff1927} G. Birkhoff, Dynamical Systems (American Mathematical Society Coil. Publ. Vol. IX, Providence, 1927), p. 82.
\bibitem{Moser1973} J. Moser, Stable and random motions in dynamical systems. Princeton University Press, 1973.
\bibitem{Verhulst1979} F. Verhulst, Discrete symmetric dynamical systems at the main resonances with applications to axi -symmetric galaxies, Philosophical Transactions of the Royal Society of London, 290, 435-465, 1979.
\bibitem{Augusteijn1980} M. F. Augusteijn and E. Breitenberger, Integration of near-resonant systems in slow-fluctuation approximation, Journal of Mathematical Physics, 21, 462—471, 1980.
\bibitem{Breitenberger1981} E. Breitenberger and R. D. Mueller, The elastic pendulum: A nonlinear paradigm, Journal of Mathematical Physics, 22, 1196-1210, 1981.
\bibitem{Ianets2017} D.Ianets and J. Shiff, Analytic Methods to Find Beating Transitions of Asymmetric Gaussian Beams in GNLS equations, Chaos, in press, 2017.
\bibitem{Hayashi2014} C. Hayashi, Nonlinear Oscillations in Physical Systems, Princeton University Press, 2014.
\bibitem{Eilbeck1985} J.C.Eilbeck, P.C. Lomdahl and A.C. Scott, The discrete self-trapping equation,  Physica D, 16, 318-338, 1985.
\bibitem{Flach2008} S. Flach, and A. Gorbach, Discrete breathers—advances in theory and applications,  Physics Reports, 467, 1 – 116, 2008.
\bibitem{Manevitch2007} L.I. Manevitch, New approach to beating phenomenon in coupled nonlinear oscillatory chains, Archive of Applied Mechanics, 77, 301-312, 2007.
\bibitem{Manevitch2014} L.I. Manevitch, A concept of limiting phase trajectories and description of highly non-stationary resonance processes, Applied Mathematical Sciences, 9, 4269-4289, 2014.
\bibitem{Manevitch2011} L.I. Manevitch, and O.V. Gendelman, Tractable modes in Solid Mechanics, Springer, Berlin, 2011.
\bibitem{Kovaleva2016} A. Kovaleva, and L.I.Manevitch, Autoresonance versus localization in weakly coupled oscillators, Physica D, 320, 1-8, 2016.
\bibitem{ManevitchRomeo} L.I.Manevitch and F. Romeo, Non-stationary resonance dynamics of weakly coupled pendula, Europhysics Letters, 112, 30005, 2015.
\bibitem{James2011} G. James, Nonlinear waves in newton's cradle and the discrete p-schrödinger equation, Mathematical Models and Methods in Applied Sciences, 21, 2335-2377, 2011.
\bibitem{Yuli2011} Y. Starosvetsky and Y. Ben-Meir, Nonstationary regimes of homogeneous Hamiltonian systems in the state of sonic vacuum, Physical Review E, 87, 062919, 2013.
\bibitem{GendelmanJSV2012} O.V. Gendelman, Analytic treatment of a system with a vibro-impact nonlinear energy sink, Journal of Sound and Vibration, Rapid Communication, 331, 4599-4608, 2012.
\bibitem{GendelmanSigalov2012} O.V.Gendelman, G. Sigalov, L.I.Manevitch, M.Mane, A.F.Vakakis and L.A.Bergman, Dynamics of an eccentric rotational nonlinear energy sink,  Journal of Applied Mechanics, Transactions of the ASME,  79, 011012, 2012.
\bibitem{LandauLifshitz} L.D.Landau and E.M.Lifshitz , Mechanics, 3 ed., Butterworth – Herrmann, 1976.
\bibitem{GoldsteinPoole} H. Goldstein, C. Poole and J. Safko,  Classical Mechanics, 3rd ed., Pearson Education International, Upper Saddle River, NJ, 2002.
\bibitem{Percival1987} I. Percival and D. Richards, Introduction to Dynamics, Cambridge University Press, Cambridge, 1987.
\bibitem{Arnold2006} V.I. Arnold, V.V. Kozlov and A.I.Neishtadt,  Mathematical Aspects of Classical and Celestial Mechanics, Springer, Berlin, 2006.
\bibitem{Itin2007} A.P.Itin, A.I. Neishtadt and A.A.Vasiliev, Captures into resonance and scattering on resonance in dynamics of a charged relativistic particle in magnetic field and electrostatic wave, Physica D, 141, 281-296, 2000.
\bibitem{Zaslavsky2007} G.M. Zaslavsky, The Physics of Chaos in Hamiltonian Systems, Imperial College Press, 2007.
\bibitem{Chirikov1979} B.V. Chirikov, A universal instability of many-dimensional oscillator systems, Physics Reports, 52, 263-379, 1979.
\bibitem{Fajans2001} J. Fajans and L. Friedland, Autoresonant (nonstationary) excitation of pendulums, Plutinos, plasmas, and other nonlinear oscillators, American Journal of Physics, 69, 1096-1102, 2001.
\bibitem{Fajans1999}  J. Fajans, E. Gilson and L. Friedland, Autoresonant (nonstationary) excitation of a collective nonlinear mode, Physics of Plasmas, 6, 4497 – 4503, 1999.
\bibitem{Vakakis2001} A.F.Vakakis and O.V. Gendelman, Energy pumping in nonlinear mechanical oscillators: Part II - Resonance capture, Journal of Applied Mechanics Transactions of ASME, 68, 42-48, 2001.
\bibitem{Sapsis2017} O.~V. Gendelman, T.~Sapsis, Energy exchange and localization in essentially nonlinear oscillatory systems: canonical formalism, Journal of Applied Mechanics 84~(1) (2017) 011009.
\bibitem{WalkerFord} G. H. Walker and J. Ford, Amplitude Instability and Ergodic Behavior for Conservative Nonlinear Oscillator Systems, Physical Review, 188, 416-432, 1969.
\bibitem{Ford1970} J. Ford and G. H. Lunsford, Stochastic Behavior of Resonant Nearly Linear Oscillator Systems in the Limit of Zero Nonlinear Coupling, Physical Review A, 1, 59-70, 1970.
\bibitem{GendelmanEscape2017} O.V.Gendelman, Escape of a harmonically forced particle from an infinite-range potential well: a transient resonance, Nonlinear Dynamics, DOI 10.1007/s11071-017-3801-x, 2017.
\bibitem{Hosea1996} M.~E. Hosea, L.~F. Shampine, Analysis and implementation of {TR}-{BDF2}, Appl. Numer. Math. 20 (1996) 21--37.
\bibitem{NPOVG2017} N. Perchikov, O.~V. Gendelman, Flat bands and compactons in mechanical lattices, Phys. Rev. E. 96 (2017) 052208.


\end{thebibliography}
\end{document}